\documentclass{aa}  
\usepackage{natbib,twoopt}
\usepackage[breaklinks=true]{hyperref} 
\bibpunct{(}{)}{;}{a}{}{,}             
\makeatletter
  \newcommandtwoopt{\citeads}[3][][]{\href{http://adsabs.harvard.edu/abs/#3}%
    {\def\hyper@linkstart##1##2{}%
     \let\hyper@linkend\@empty\citealp[#1][#2]{#3}}}
  \newcommandtwoopt{\citepads}[3][][]{\href{http://adsabs.harvard.edu/abs/#3}%
    {\def\hyper@linkstart##1##2{}%
     \let\hyper@linkend\@empty\citep[#1][#2]{#3}}}
  \newcommandtwoopt{\citetads}[3][][]{\href{http://adsabs.harvard.edu/abs/#3}%
    {\def\hyper@linkstart##1##2{}%
     \let\hyper@linkend\@empty\citet[#1][#2]{#3}}}
  \newcommandtwoopt{\citeyearads}[3][][]%
    {\href{http://adsabs.harvard.edu/abs/#3}
    {\def\hyper@linkstart##1##2{}%
     \let\hyper@linkend\@empty\citeyear[#1][#2]{#3}}}
\makeatother

\usepackage{adjustbox}
\usepackage{graphicx}
\usepackage{txfonts}
\usepackage{longtable}
\usepackage{tabularx, blindtext}
\usepackage{rotating}
\usepackage{pdflscape}
\usepackage{inputenc}
\usepackage{caption}
\DeclareCaptionFormat{continued}{#1 (continued)#2#3\par}

\begin{document} 
\title{Physical characterization of recently discovered globular clusters in the Sagittarius dwarf spheroidal galaxy}
\subtitle{II. Metallicities, ages and luminosities}
   \author{E.R. Garro \inst{1}
          \and
          D. Minniti\inst{1,2}
            \and
         M. Gómez\inst{1}
          \and
           J. Alonso-García\inst{3,4}        
             }
   \institute{Departamento de Ciencias Físicas, Facultad de Ciencias Exactas, Universidad Andres Bello, Fernández Concha 700, Las Condes, Santiago, Chile
   \and
 Vatican Observatory, Vatican City State, V-00120, Italy
  \and
 Centro de Astronomía (CITEVA), Universidad de Antofagasta, Av. Angamos 601, Antofagasta, Chile
 \and
 Millennium Institute of Astrophysics , Nuncio Monse\~nor Sotero Sanz 100, Of. 104, Providencia, Santiago, Chile
}
  \date{Received April 13, 2021; Accepted July 21, 2021}

 
  \abstract
   {
Globular clusters (GCs) are important tools to rebuild the accretion history of a galaxy. In particular, there are newly discovered GCs in the Sagittarius (Sgr) dwarf galaxy, that can be used as probes of the accretion event onto the Milky Way (MW). }
   {
    Our main aim is to characterize the GC system of the Sgr dwarf galaxy by measuring its main physical parameters.
   }
   {
We build the optical and near-infrared color-magnitude diagrams (CMDs) for 21 new Sgr GCs using the VISTA Variables in the Via Lactea Extended Survey (VVVX) near-infrared database combined with the Gaia Early Data Release 3 (EDR3) optical database. We derive metallicities and ages for all targets, using the isochrone-fitting method with PARSEC isochrones. We also use the relation between RGB-slope and metallicity as an independent method to confirm our metallicity estimates. In addition, the total luminosities are calculated both in the near-infrared and in the optical. We then construct the metallicity distribution (MD), the globular cluster luminosity function (GCLF),  and the age-metallicity relation for the Sgr GC system.
   }
   {
We find that there are 17 metal-rich GCs with $-0.9 < [Fe/H] < -0.3$, plus 4 metal-poor GCs with $-2.0 < [Fe/H] < -1.1$ in the new Sgr GC sample. There is a good agreement between the metallicity estimates using isochrones and RGB slopes. Even though our age estimates are rough, we find that the metal-poor GCs are consistent with an old population with an average age of $\sim$13 Gyr, while the metal-rich GCs show a wider age range, between $6 - 8$ Gyr and $10 - 13$ Gyr. Additionally, we compare the MD and the GCLF for the Sgr GC system with those of the MW, M31 and Large Magellanic Cloud (LMC) galaxies.}
{
  We conclude that the majority of the metal-rich GCs are located within the main body of Sgr galaxy. We confirm that the GCLF is not a universal distribution, since the Sgr GCLF peaks at fainter luminosities ($M_V \approx -5.5$ mag) than the GCLFs of the MW, M31 and LMC. Also, the MD shows a double-peaked distribution, and we note that the metal-rich population looks like the MW bulge GCs. We compared our results with the literature concluding that the Sgr progenitor could have been a reasonably large galaxy able to retain the supernovae ejecta, thus enriching its interstellar medium.}
   \keywords{Galaxies: dwarf – Galaxy: halos -- Galaxies: luminosity function, mass function -- Galaxy: stellar content – (Galaxy:) globular clusters: general – Infrared: stars – Surveys}

\titlerunning {Physical characterization for recently discovered globular clusters in the Sagittarius dwarf galaxy}
\authorrunning {E.R. Garro et al.}
   \maketitle

\section{Introduction}
The main accepted galaxy formation paradigm predicts that galaxies grow hierarchically through mergers with other galaxies (e.g., \citealt{SZ1978,WR1978}), and thus the accretion of diffuse gas and dark matter occur especially into the halo. There are different types of mergers, but broadly we can distinguish major mergers from minor mergers, depending on the mass ratio of the two objects. In the first case,  the masses of the two colliding galaxies are comparable, while in the second case a galaxy of lower mass is accreted into a more massive galaxy. Obviously,  when a galaxy is observed,  evidence of substructures and deviations from symmetry are indicative of past or/and ongoing mergers. Clues can be detected within galaxies themselves in different forms, such as streams,  bridges, density waves, overdensities of stars, substructure in galaxy's gas,  different globular cluster populations as well as changes in kinematics. Proofs of merging events are found both in the Galaxy \citep{Newberg_2002,Belokurov_2006,Martin_2014} and in external galaxies (i.e., \citealt{Einasto2012,Cohen_2014, Abraham_2018}),  like e.g.  the Andromeda galaxy (M31,  \citealt{Ibata2001}).  Therefore, the history of our Milky Way is a history of accretion.  At least seven past accretion events can be singled out: Kraken \citep{Kruijssen2019,Kruijssen2020},  Sequoia, \citep{Myeong2019} Sagittarius \citep{Ibata1994}, Helmi stream \citep{Helmi1999} and \textit{Gaia}-Enceladus \citep{Helmi2018},  and both Large and Small Magellanic Clouds (LMC and SMC) will infall towards the MW,  we know that the Magellanic system is likely on its first passage about the Milky Way \citep{Kallivayalil_2006,Besla_2010, Kallivayalil_2013}.\\

The most representative example of a satellite galaxy being accreted by our Milky Way is the Sagittarius dwarf spheroidal galaxy (Sgr dSph).  Discovered by \cite{Ibata1994},  it is located behind the Galactic bulge at heliocentric distance $D\approx 26.5$ kpc \citep{Monaco2004, Hamanowicz2016,Vasiliev2020} and at about 6.5 kpc below the Galactic plane. The Sgr dSph represents an excellent laboratory, since the tidal destruction process is still ongoing \citep{Majewski2003,Law_2010, Belokurov2014}.  The infall into the MW has been estimated to occur $8\pm 1.5$ Gyr ago by \cite{Dierickx_2017} and $9.3 \pm 1.8$ Gyr ago by \cite{Hughes2019}.  However, many questions about its formation and evolution before and after its accretion inside the Galactic halo still remain unanswered.  

The mass of the Sgr progenitor and the present-day mass of the remnant are still topics of active discussion. The stellar mass of the main body is $M_{\ast}\sim 2\times 10^{7}\ M_{\odot}$ \citep{Ibata2004} and the dynamical mass is $M_{dyn}\sim 2\times 10^{8}\ M_{\odot}$ \citep{Grcevich2009}. However, the subsequent census of its stellar content revealed that its total mass could be as high as $10^{11}\ M_{\odot}$ including its dark matter halo \citep{Niederste-Ostholt2012, Laporte2018,Vasiliev2020},  indicating that this was a major merging process between our Galaxy and the Sgr dSph.  Nevertheless, the accretion event of the progenitor of the Sgr dwarf was a minor merger with mass ratio of $1:104^{+70} _{-43}$. Indeed, the accretion has occurred at $z<1$, when already the MW was completely formed and its stellar mass was $M > 10^{10} M_{\odot} $ (\citealt{Kruijssen2020}; see their Fig. 9).\\

Although this satellite appears to be quite elongated out to $\sim 100$ kpc \citep{Majewski2003, Law_2010}, its main body contains an overdensity of stars, which is concentrated in its centre, where the massive and metal-poor globular cluster NGC~6715 (M~54) is located.  It is also coincident in position with the nucleus of the dwarf galaxy (e.g., \citealt{Bassino1995,Layden2000}),  although \cite{Bellazzini2008} have argued, from measurements of velocity dispersion profiles, that M~54 is not the core of Sgr but instead it may have formed independently and plunged to the core of Sagittarius due to dynamical friction.  Although a large number of star clusters would be expected in these regions, there are only 9 known and well-characterized GCs associated with Sgr. Besides NGC~6715,  there are Arp~2,  Terzan~7,  and Terzan~8 located in the main body, and Palomar~12, Whiting~1,  NGC~2419,  NGC~4147, and NGC~5634 situated in the extended tidal streams \citep{Bellazzini2020}.  We may expect that stars and globular clusters in the stream exhibit a different age-metallicity relation (AMR) to those formed in the central galaxy \citep{Forbes_Bridges2010,Leaman2013,Kruijssen2019}. Indeed, the stellar population seems to be divided into three groups. It is dominated by a metal-rich and intermediate-age ($[Fe/H]= -0.4$ to $-0.7$~dex,  $t= 5$ to $8$ Gyr;  e.g., \citealt{Layden2000,Bellazzini2006a}) population in the central part of the galaxy.  Additionally, hints of a young metal-rich population ($[Fe/H]=-0.4$ and $t=2.5$ Gyr) have been found, including also stars of solar-abundance (e.g., \citealt{Monaco2005b, Chou2007}). On the other hand, there is also an old and metal-poor ($t=10$ to $13$ Gyr, $[Fe/H]\sim -2.2$; \citealt{Momany2005}) counterpart.  \\
In addition, \cite{Mucciarelli2017},  analysing 235 giant stars, detected a metallicity gradient within the Sgr nucleus. They found two peaks of the metal-rich population,  indicating that the stars in the metal-rich component formed outside in over a few Gyr with $ [Fe/H]=-0.58$ ($38 \leq R\leq 70$ kpc), with each subsequent generation of stars more centrally concentrated with $[Fe/H]=-0.38$ ($R\leq 19$ kpc). They also mentioned that some memory of its formation may still be detectable since the stellar population is not dynamically mixed.  Also,  they suggested that Sgr was affected by a strong gas loss occurring 7.5 Gyr to 2.5 Gyr ago, presumably starting at the first peri-Galactic passage of the dwarf after its infall into the Milky Way. \\

In all cases, it is crucial to complete the census of the GC system in the Sgr dSph in order to put formation and evolution constrains. The first step was done by \cite{Minniti2021_first},  who identified 23 GC candidates that may belong to the Sgr dSph, using the VVVX survey,  and already confirm from their further analysis 12 of them as bona-fide members.  Additionally,  \cite{Minniti2021_second} later discovered 18 more GCs which might belong to the Sgr galaxy,  confirming 8 of them already as bona-fide members.  In both papers (hereafter M21, for simplicity), physical parameters such as reddening, extinction, and distance for each cluster have been estimated.  In this follow-up paper, we calculate other important parameters: metallicities, luminosities and, where possible, we give a rough estimate of the ages, in order to explore the AMR.  \\

In Section 2, we briefly describe the observational data.  In Section 3, we explain the methods used to estimate the physical parameters and the resulting values for each Sgr GC.  In Section 4,  we show the metallicity distribution (MD) and luminosity function (LF) for all the GCs in the Sgr dSph, including both new detections from M21 and previously known ones,  and we compare these distributions with the MD and LF of the MW, M31,  and LMC.  In Section 5, a summary and conclusions are given.

\section{Observational datasets: VVVX, 2MASS and Gaia EDR3}
We use the deep near-infrared (IR) data from the VISTA Variables in the Via Lactea (VVV; \citealt{Minniti2010, Saito2012}) and its eXtention (VVVX; \citealt{Minniti2018}) surveys,  acquired with the VISTA InfraRed CAMera (VIRCAM) at the 4.1m wide-field Visible and Infrared Survey Telescope for Astronomy (VISTA; \citealt{Emerson2010}) at ESO Paranal Observatory.  Both VVV and VVVX data are reduced at the Cambridge Astronomical Survey Unit (CASU; \citealt{Irwin2004}) and further processing and archiving is performed with the VISTA Data Flow System (VDFS; \citealt{Cross2012}) by the Wide-Field Astronomy Unit and made available at the VISTA Science Archive and ESO Archive. 
In our analysis, we use a preliminar version of the VVVX photometric catalogue (Alonso-Garc\'ia et al, in prep.), which extracts the point-spread function (PSF) photometry from the VDFS-reduced images. To build this photometric catalogue, a similar analysis to that described in \cite{AlonsoGarcia2018} for the VVV original footprint was followed. In order to increase the dynamic range of  our near-IR photometry, we merged\footnote{To merge the catalogs, first we transform the VVVX catalogs, which are in the VISTA magnitude system, into the 2MASS magnitude scale by applying the recipee from http://casu.ast.cam.ac.uk/surveys-projects/vista/technical/photometric-properties .} our deep VVVX photometry with the fainter catalogues from the Two Micron All Sky Survey (2MASS; \citealt{Skrutskie2006}). We are able to provide in this way accurate photometry also for bright stars ($K_s<11$ mag) that are saturated in the VVVX images. We also use the $K_s$-band photometry from \cite{McDonald2013} and \cite{McDonald2014} for only Minni327 that is located outside the VVVX area. \\
On the other hand, we use the recent optical photometry from the \textit{Gaia} Early Data Release 3 (EDR3) \citep{GaiaCollaboration2020} in order to take advantage from the more precise astrometry and proper motions (PM), which were employed especially in the first part of the work (M21) to discriminate the nature of the candidates and the inclusion to the Sgr galaxy.  In this work, by matching the \textit{Gaia} EDR3 and VVVX datasets,  we construct optical and near-IR colour-magnitude diagrams (CMDs) in order to obtain the metallicity, age and luminosity for each target. 

\section{Estimation of parameters for the new Sgr GCs}
We focus on the 21 GCs that are recognized as confirmed Sgr members in M21. We list them and summarize their main physical properties in Table \ref{tablegcs}.\\
Along the line of sight to the Sgr dwarf galaxy,  high contamination from the nearby Milky Way disk and from the more distant bulge field stars may represent an obstacle,  however M21 applied a PM decontamination procedure that allows us to work on clean catalogues.  Summarizing, M21 performed two tests in order to estimate the statistical significance of the stellar overdensities in the Sgr main body. First, following the procedure of \cite{Koposov2007}, they calculated the number of stars in excess with respect to the background field, whose random fluctuations are assumed to be Poissonian. After that, they compared that excess with the statistical error on the background number counts, and finally they revealed the cluster detection if the significance is larger than $3\sigma$. Whereas, the second method concerns the variation of the background. In this case, M21 computed the number counts of sources included within several adjacent circles ($r < 3'$) around a wider area from each cluster centre coordinates. Subsequently, they calculated the standard deviation of the distribution of these number counts to derive the signal-to-noise, and considered as significant all candidate GCs with detections larger than $3\sigma$. \\ 

\begin{figure*}
\centering
\includegraphics[width=4cm, height=4cm]{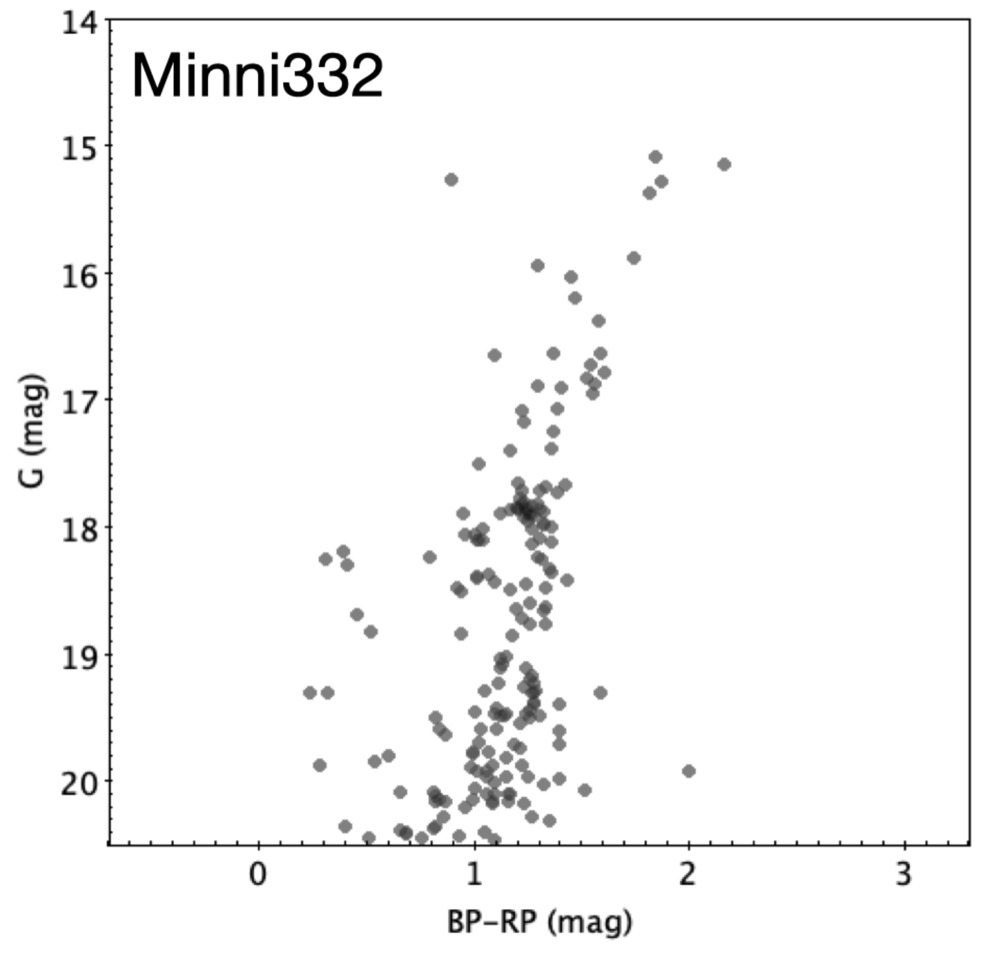} \qquad \quad
\includegraphics[width=5cm, height=4cm]{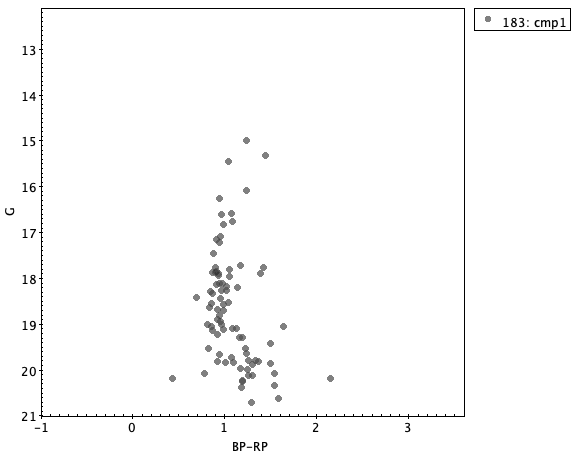}
\includegraphics[width=5cm, height=4cm]{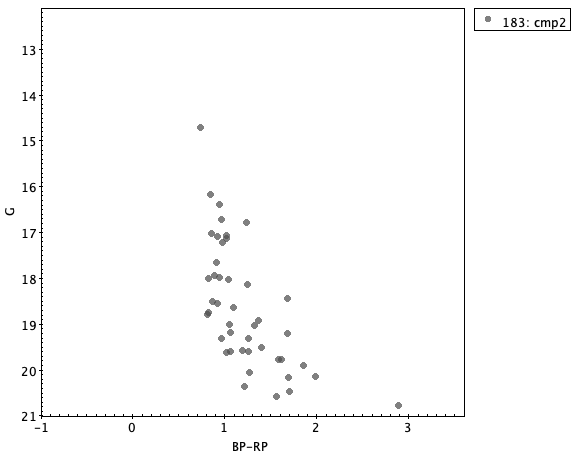}
\includegraphics[width=5cm, height=4cm]{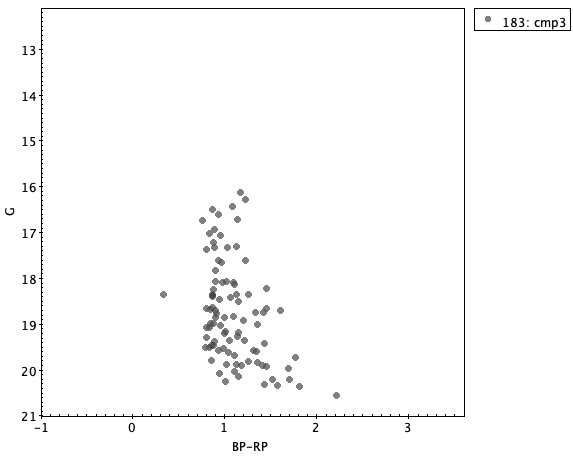}
\includegraphics[width=5cm, height=4cm]{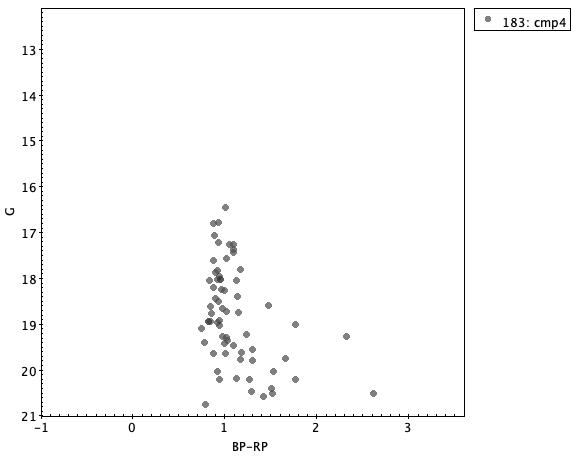}
\includegraphics[width=5cm, height=4cm]{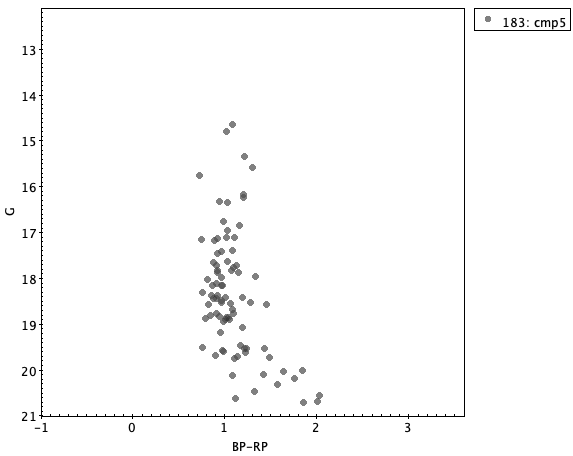}
\caption{Comparison between the Gaia EDR3 optical CMDs of Minni332, taken as a representative of the Sgr GCs \citep{Minniti2021_second},  and five bulge fields located at similar latitudes but away from the Sgr main body.} 
\label{sgr_bulge_comp}
\end{figure*}

Our main goals are to derive reliable values of metallicity, age and luminosity in order to build up an "updated" LF as well as a MD for this satellite galaxy.  \\
First, we have built the optical and near-IR CMDs for all the GCs in our sample (Fig. \ref{cmds}),  confirming that our targets are not bulge GCs,  which should be $\sim 2.5$ magnitudes brighter in the mean, as demonstrated by previous works \citep{Minniti2010,Minniti_2017a,Minniti_2017b, Minniti_2018, Palma2019,Garro2021a}. Additionally, we have made a comparison between the Sgr GCs and bulge fields CMDs.  Summarizing,  we first selected five bulge fields at the same latitude but $\sim 7^{\circ}$ away from the Sgr main body.  As done for our objects, we constructed the \textit{Gaia}-VVVX catalogues,  properly decontaminated applying parallax and PM cuts.  Figure \ref{sgr_bulge_comp} shows the comparison between these five bulge fields and Minni332 (taken as representative of the Sgr GCs) CMDs. This proves that Sgr GCs, shown in Fig. \ref{cmds}, are not affected by the bulge population, and obviously it demonstrates that the Sgr GCs here investigated are not bulge GCs. \\

Following the same strategy as in \cite{Garro2020,Garro2021a},  we prefer the isochrone-fitting method in order to obtain robust results.  We favour the PARSEC alpha-enriched isochrones \citep{Bressan2012,Marigo2017}, and adopt the reddening, extinction and distance modulus values, previously calculated by M21.
Briefly,  M21 derived appropriate reddening corrections using the maps of \cite{Schlafly2011} and adopting the following relations for the extinctions and reddenings: $A_{Ks}=0.11 \times A_{V}$,  $A_{Ks}=0.72 \times E(J-K_s)$,  $A_G=0.79 \times A_{V}$ and $A_G=2.0 \times E(BP-RP)$, obtaining a fairly uniform reddening. Additionally, M21 employed the RC absolute magnitudes and intrinsic colour by \cite{RuizDern2018}, in order to compute distances for each individual cluster.
Thereafter, we fit in particular the position of RC,  BHB and brighter stars.  The best fitting age and metallicity was obtained by comparing our data with isochrones generated with different ages and metallicities and selecting the best by-eye fit.  We first fixed the age and varied the metallicity. Then we fixed the metallicity and searched for the best fitting age.  The variations of the age and metallicity have different effects along the evolutionary stages.  Especially,  changing  metallicity strongly affects the RGB and the RC,  conversely changes in ages modify the position of the MS-TO regions.  In this way, we assign an error to age and metallicity,  changing simultaneously both the metallicity and age values until the fitting-isochrones did not reproduce all evolutionary sequences in both optical and near-IR CMDs. 

We find 17 metal-rich GCs, specifically ten with $[Fe/H]=-0.3$ to $-0.5$ dex,  and seven with $[Fe/H]$ between $-0.6$ and $-0.9$ dex, whereas other four GCs are metal-poor with $[Fe/H]=-1.1$ to $-2.0$ dex.  Indeed,  inspecting the luminosity function for each GC (Fig.  \ref{LF}), we single out the RC  as a peak in the histogram,  confirming that the stellar population is metal-rich. Giving some examples, we can appreciate a clear excess in Minni324, 330, 340, 342, 344 histograms.  Also,  metal-poor clusters show blue horizontal branch stars (BHBs) as well as RR Lyrae variable stars (Section 3.1),  e.g. Minni01,  147,  and 335. \\

Using the same method, we have tried to derive the age, but the absolute estimate of this parameter results to be a challenge when the magnitude of the main sequence turn-off (MSTO) is below the detection limit.  However, there are a variety of works (e.g., \citealt{Rosenberg_1999,Gratton2003}), adopting relative age determination (instead of absolute ages) from the observable CMDs, although less accurate. 
We can therefore obtain a solid lower limit for the age because the differences $\Delta K_s$(HB-MSTO) $>1.80$ mag and $\Delta G$(HB-MSTO) $>3$ mag, meaning that the clusters are not young with an $Age \gtrsim 7$ Gyr.  This could be improved upon by measuring the extension of the giant branch, as the presence of bright and red stars in an extended AGB is indicative of an intermediate-age system (e.g.,  \citealt{Freedman2020}, and references therein).  Even so,  it is hard to distinguish AGB stars from RGB stars in such sparse CMDs.  In addition, for the clusters containing RR Lyrae a more stringent age limit can be derived from the typical ages of these variable stars, that are older than about 10 Gyr see Section 3.1).

Taking into account the lower limit found for the age and the RR Lyrae cluster membership,  but also comparing our clusters with known Sgr GCs and their observable CMDs,  we obtain more constrained ages for Minni 01, 146, 148, 147, 148, 311,  326, 330, 335, 338 and 343, finding all of them to be old GCs. 
Additionally, we note that Minni 144, 145, 324, 325, 328, 329, 332, 340, 341, 342, and 344 are intermediate/old globular clusters with an age $t > 8$~Gyr. Hence, deeper observations are necessary to better constrain the ages of these GCs.  \\

It is well-known that the position and the morphology of the RGB in the CMD strongly depend on the metal content of the stellar population: the higher the metal content, the cooler the effective
temperature $T_{eff}$ and the redder the RGB stars (e.g., \citealt{Ferraro_2000,Valenti2005}).  Therefore,  a series of empirical parameters (i.e.,  RGB colours at fixed level of magnitude, RGB magnitude at fixed colour, and the RGB slope) can be used to derive a photometric estimate of the global metallicity of the considered stellar population.  Indeed, we preferred to use the slope$_{RGB}$ as another independent-method to constrain our metallicity estimates. Specifically, we adopted the $slope_{RGB}-[Fe/H]$ linear relation by \cite{Cohen_2015} in the near-IR passband.   We have derived the RGB slope as the line connecting two points along the RGB at the HB level and $\sim $2.5 mag brighter. The resulting values, listed in Table \ref{tablegcs} (column 9 and 10), are in very good agreement to those derived through isochrone-fitting. \\ 

Finally, we estimate the total luminosity $M_{K_s}$ of all the GCs in our sample.  Certainly,  the total luminosity varies from GC to GC, as it depends on several factors, such as the luminosity of the brightest stars, the distribution of members along the isochrones and the presence of red giants.  We first measured the flux for each star and consequently the total flux for the target, and derived the absolute magnitude converting the total flux. All resulting $M_{Ks}$ are listed in Table \ref{tablegcs}.  We also derive the equivalent absolute magnitude in $V-$band, assuming that the mean $(V-K_s)=2.5\pm 0.5$ mag for observed GCs in systems like the MW and M31 (e.g.,  \citealt{Barmby2000,Cohen2007,Conroy_2010}).  Although this could be a rough approximation,  studies (e.g., \citealt{Pessev2008}) have demonstrated that the smallest spread in intrinsic colours is found for clusters with ages $\gtrsim 10$ Gyr, whereas the larger spread in colour is found for clusters in the age range $2-4$ Gyr.   
On the other hand, comparing the intrinsic colour with M31 GC system,  \cite{Wang_2014} found a good correlation between $(V-K_s)$ and metallicity (with a mean value $\sim 2.5$),  even if the relation shows a notable departure from linearity with a shallower slope toward the redder end ($[Fe/H]\gtrsim -0.1$).  Additionally, the $V-$band total luminosities should be trusted only to $1-1.5$ mag, which is the scatter in these mean integrated colour. Hence,  the use of a mean $(V-K_s)=2.5$ is an acceptable compromise for our targets.\\
These luminosities,  calculated in this way,  are an underestimate of the total luminosity since the faintest stars are missing.  Even though \cite{Kharchenko2016}, analysing MW star clusters, noted that the cluster luminosity profiles show similar features, such as a relatively fast rise in the luminosity at small $\Delta K_s$\footnote{\cite{Kharchenko2016} defined $\Delta K_s$ as the magnitude difference between a given cluster member and the brightest cluster member. } owing to the dominant contribution of a few bright stars and a much slower increase by successively including fainter stars.  As a rule, the first 10--12 brighter members accumulate more than half of the integrated luminosity of  a cluster, which no longer changes at $\Delta K_s>10$. \\
Consequently,  we derived the total luminosity in $V-$band ($M_{Vtot}$), comparing our GC luminosities with other known GCs with similar metallicity,  in order to estimate the fraction of luminosity that comes from low-mass stars.  We explain the main steps computed to estimate the total luminosity for each Sgr GCs in the Appendix A. \\
We find that all new Sgr GCs are low-luminosity objects,  at least $\sim 1.3 $ mag less luminous than the MW GC luminosity function peak ($M_V = (-7.4 \pm 0.2)$ mag from \cite{Harris_1991, ashman_1998}). \\

Once obtained the absolute magnitudes in $K_s -$ and $V-$bands,  we are able to estimate their masses.  We first assume a typical GC mass-to-light ratio $M/L_{K} \approx 1$, equivalent to $M/L_V \approx 2$  \citep{Haghi2017,Baumgardt2020}.  Subsequently,  since \cite{Haghi2017} exhibited the $M/L - $[Fe/H] relations (see their Fig.2),  we derive the $M/L$ values, depending on our metallicity estimates, in order to obtain more rigorous results.  In both cases, we find low-mass GCs,  with the same order of magnitude $M \approx10^3$ to $10^4\ M_{\odot}$ (see Table \ref{tablegcs}).  Additionally, we calculate the mass for each known Sgr GC listed in Table \ref{tablecomp},  finding a good agreement between our values and those listed in the Galactic Globular Cluster Database version 2\footnote{https://people.smp.uq.edu.au/HolgerBaumgardt/globular/} by \cite{Hilker2020}. Clearly,  the difference are due to the large scatter in the Haghi's relations ($\Delta(M/L_K)\approx \pm 0.5$ and $\Delta(M/L_V)\approx \pm 1.0$).

All these parameters are summarized in Table \ref{tablegcs}. We also marked with an asterisk symbol all GCs that do not show well-populated RGBs, as they could have less accurate parameters.  However, we include them in our analysis because they have no effect on the final results.
\begin{table*}
\centering 
 \begin{adjustbox}{max width=\textwidth}
\begin{tabular}{lccccccccccc}
\hline\hline
Cluster ID & L      &  B        &      $[Fe/H]_{iso}$             & $M_{Ks} $     & $M_{Vtot}$        &   Age\tablefootmark{a}  &  $slope_{RGB}$ & $[Fe/H]_{slope}$\tablefootmark{b} &$M/L_{K}$\tablefootmark{c} & M\tablefootmark{c}& $S_{RR}$ \\
                  & [deg] & [deg] & [dex] & [mag]  &[mag] & [Gyr]&     & [dex] &  & [$M_{\odot}$]&\\
\hline
Minni01$^{\ast}$  & 5.3706 &  -9.3482  &   $-1.2  \pm  0.2$  &  $-5.8 \pm 0.8$  & $ -3.4 $    &   $[12] 10-13$  &  -0.08   &  -1.22 &0.9   & $3.8 \times 10^{3}$ & 43.65\\
Minni144  & 4.1693 &  -11.1990  &   $-0.9  \pm  0.2$  &  $-7.0 \pm 0.7$ & $-5.1$ &   $[11] 10-13$    &  -0.09   &  -0.90 &0.8   & $1.0 \times 10^{4}$ & 9.12\\
Minni145  & 7.2695 &  -12.9988 &   $-0.8 \pm  0.3$  &  $-6.8 \pm 1.1$  &  $-4.4 $& $[10] 10-13  $   &  -0.092   &  -0.85 &0.8   & $8.5 \times 10^{3}$ & 86.89\\
Minni146$^{\ast}$  & 3.9698 &  -14.1097 &   $-1.1 \pm  0.2$  &  $-6.9 \pm 1.1$  &   $-4.5 $ &$[12] 7-13   $    &  -0.084   &  -1.09  &0.9   & $1.1 \times 10^{4}$ & 0\\
Minni147  & 3.9993 &  -11.6982 &   $-2.0  \pm  0.2$  &  $-6.7 \pm 1.0$  &$-4.4$&   $[13] 10-13  $    &  -0.07   &  -1.53   &1.1   & $1.1 \times 10^{4}$ & 17.38\\
Minni148  & 5.3598 &  -13.3792 &   $-0.3  \pm  0.2$  &  $-7.2 \pm 1.1$  & $-5.4$&  $[10] 10- 13$    &  -0.11   &  -0.30  &0.4   & $6.2 \times 10^{3}$ & 20.75\\
Minni311$^{\ast}$  & 5.2749 &  -9.32097 &   $-0.6  \pm  0.3$  &  $-8.0 \pm 1.1$  & $-5.7$&  $[10] 7-13   $    &  -0.09   &  -0.90  &0.7   & $2.3 \times 10^{4}$ & 0\\
Minni324  & 5.7821 &  -11.9392 &   $-0.5  \pm  0.2$  &  $-6.9 \pm 1.1$  &  $-5.0$& $[13]10-13 $    &  -0.10   &  -0.60 &0.5   & $5.8 \times 10^{3}$& 20.0\\
Minni325$^{\ast}$  & 4.1172 &  -14.5024 &   $-0.4  \pm  0.2$  &  $-6.9 \pm 1.3$ &  $-5.1$& $[10] 7-13  $    &  -0.108  &  -0.36 &0.4   & $4.7 \times 10^{3}$ & 0\\
Minni326  & 5.7635 &  -13.0909 &   $-0.3  \pm  0.2$  &  $-7.3 \pm 1.2$  & $-5.1$ &   $[10]10 - 13$  &  -0.11   &  -0.30  &0.4   & $6.8 \times 10^{3}$  & 18.93\\
Minni328  & 5.2657 &  -12.3993 &   $-0.5  \pm  0.1$  &  $-6.9 \pm 0.9$  &  $-5.1$&  $[10]10-13 $    &  -0.10	  & -0.60 &0.5   & $5.8 \times 10^{3}$& 9.12\\
Minni329  & 5.0863 &  -11.8213 &   $-0.7  \pm  0.2$  &  $-7.4 \pm 1.3$  & $ -5.3$& $[13]7-13$     &  -0.096  &  -0.73 &0.8   & $1.5 \times 10^{4}$ & 0 \\
Minni330  & 6.1214 &  -14.0065 &   $-0.6  \pm  0.2$  &  $-7.2 \pm 1.1$ & $-4.9$&   $[10]10 - 13$     &  -0.096   &  -0.73&0.7   & $1.1 \times 10^{4}$&10.96\\
Minni332  & 5.9722 &  -12.1903 &   $-0.5  \pm  0.2$  &  $-7.0\pm 1.1$  & $-5.1 $&   $[11]10-13$    &  -0.101  &  -0.58  &0.5   & $6.4 \times 10^{3}$ & 9.12\\
Minni335  & 5.1161 &  -12.4645 &   $-1.3  \pm  0.3$  &  $-7.7\pm 1.0$  &  $-5.5$ & $[14] 10-14  $    &  -0.076  &  -1.33 &1.0   & $2.4 \times 10^{4}$&25.24\\
Minni338  & 4.4542 &  -14.5275 &   $-0.6  \pm  0.1$  &  $-7.6\pm 1.1$  & $-5.3$ &   $[10]7-13   $    &  -0.096  &  -0.73 &0.7   & $1.6 \times 10^{4}$ &0\\
Minni340  & 5.2576 &  -13.2629 &   $-0.5  \pm  0.2$  &  $-8.0\pm 1.1$  &  $-6.1$& $[10]10-13$    &  -0.104  &  -0.48&0.5   & $1.6 \times 10^{4}$ &7.26\\
Minni341  & 5.4627 &  -13.8907 &   $-0.5  \pm  0.2$  &  $-8.0 \pm 1.2$  & $-6.1$&  $[10]10-13 $       &  -0.104  &  -0.48&0.5   & $1.6 \times 10^{4}$ & 7.26\\
Minni342  & 4.9365 &  -14.1964 &   $-0.5  \pm  0.2$  &  $-7.4 \pm 1.1$  & $-5.5$   &$[10]10-13$   &  -0.099  &  -0.64 &0.5   & $9.3 \times 10^{3}$ &  6.31\\
Minni343  & 5.0578 &  -14.4441 &   $-0.8  \pm  0.2$  &  $-7.8 \pm 1.1$  & $-5.4$&  $[13] 10-13   $   &  -0.092  &  -0.85&0.8   & $2.2\times 10^{4}$  & 20.75\\
Minni344  & 5.1483 &  -14.6487 &   $-0.4  \pm  0.2$  &  $-8.3 \pm 1.3$  & $-6.0$&   $[10]10-13  $    &  -0.104  &  -0.48 &0.4   & $1.7 \times 10^{4}$& 3.98\\
\hline\hline
\end{tabular}
\end{adjustbox}
\caption{Position, metallicity, age, luminosity,  mass-to-light ratio,  mass and specific frequency of RR Lyrae stars for all new Sgr GCs, analysed in this work.  We marked with asterisk symbol the GCs that do not show well-populated CMDs, thus they could have less reliable parameters.  We also marked with an asterisk symbol all GCs that do not show well-populated CMDs, as they could have less accurate parameters. } 
\tablefoot{
   \tablefoottext{a}{We highlight the age used in the fit of the isochrones (Fig. \ref{cmds})  in square brackets.}
   \tablefoottext{b}{Calculated using the $slope_{RGB}-[Fe / H]$ relation by \cite{Cohen_2015}.}
   \tablefoottext{c}{Mass estimated by the present work, adopting different values of $M/L_{K}$ from \cite{Haghi2017}, depending on our metallicity values.}
   }
\label{tablegcs}
\end{table*}

\begin{table*}
\centering 
\begin{tabular}{lcccccccc}
\hline\hline
Cluster ID &  L        & B         &    $[Fe/H]$  &    $M_{Ks}$&      $M_{V}$&    Age  & M\tablefootmark{a} & M\tablefootmark{b}  \\
                  & [deg] & [deg]  & [dex] & [mag] & [mag] & [Gyr] & $[M_{\odot}]$ \\
\hline
NGC 6715    &  5.6070   & -14.0871  &   -1.49 & -12.51&   -9.98&  13.0 & $1.62\times 10^{6}$ & $2.3 \times 10^{6} $  \\
Terzan 8       &  5.7592   & -24.5587  &   -2.16 & -7.55 &   -5.07&  13.0 & $5.8\times 10^{4}$& $2.6 \times 10^{4} $  \\
Arp 2       &  8.5453   & -20.7853  &   -1.75 & -7.79 &   -5.29&  11.3 & $3.8\times 10^{4}$  & $3.2 \times 10^{4} $\\
Terzan 7       &  3.3868   & -20.0665  &   -0.32 & -7.55 &   -5.01&   7.5& $2.0\times 10^{4}$& $8.6 \times 10^{3} $   \\
Palomar 12      &  30.5101  & -47.6816  &   -0.85 & -6.98 &   -4.48&   9.0 & $6.4 \times 10^{3}$ & $1.0 \times 10^{4} $ \\
Whiting 1   & 161.6160  & -60.6363  &   -0.7  & -4.96 &   -2.46&   6.5 & $1.6\times 10^{3}$ & $1.6 \times 10^{3} $ \\
NGC 2419    & 180.3697  &  25.2417  &   -2.15 & -11.92&   -9.42&   12.3 & $1.4\times 10^{6}$ & $1.4 \times 10^{6} $  \\
NGC 4147    & 252.8483  &  77.1887  &   -1.84 & -8.67 &   -6.17&   14.0& $3.8\times 10^{4}$& $6.6 \times 10^{4} $  \\
NGC 5634    & 342.2093  &  49.2604  &   -1.88 & -10.19&   -7.69&   13.0& $2.2\times 10^{5}$ & $2.9 \times 10^{5} $ \\
\hline\hline
\end{tabular}
\caption{Position, metallicity, age, luminosity and mass for the previously known Sgr GCs, used for comparison.}
\tablefoot{
   \tablefoottext{a}{Mass values by the Galactic Globular Cluster Database version 2 \citep{Hilker2020}}
   \tablefoottext{b}{Mass estimated by the present work, adopting different values of $M/L_{K}$ from \cite{Haghi2017}, depending on our metallicity values.}
   }
\label{tablecomp}
\end{table*}

\section{RR Lyrae stars in Sgr GC system}
RR Lyrae stars are usually excellent tracers of metal-poor and old populations in the MW.  M21 searched for these stars within $3'$ and $10'$ from the cluster centres.  As expected, they found an excess of RR Lyrae in some metal-poor GCs, such as in Minni01 ($N_{3'}=1$; $N_{10'}=3$), Minni147 ($N_{3'}=1$; $N_{10'}=6$) and Minni335 ($N_{3'}=4$; $N_{10'}=12$). We should suppose the presence of these variable stars also in GCs with intermediate metallicity,  like in Minni144 ($N_{3'}=1$; $N_{10'}=7$),  Minni145 ($N_{3'}=5$; $N_{10'}=14$) and Minni343 ($N_{3'}=3$; $N_{10'}=13$). While, depending on the traditional stellar evolutionary theory, we do not expect the presence of many RR Lyrae stars in metal-rich GCs.  However, a few metal-rich GCs in the Sgr dwarf show a higher number of RR Lyrae within $10'$ from the cluster centre: Minni148 ($N_{3'}=3$; $N_{10'}=20$), Minni324 ($N_{3'}=2$; $N_{10'}=9$), Minni326 ($N_{3'}=3$; $N_{10'}=17$),  Minni328 ($N_{3'}=1$; $N_{10'}=9$), Minni330 ($N_{3'}=1$; $N_{10'}=11$), Minni332 ($N_{3'}=1$; $N_{10'}=7$),  Minni340 ($N_{3'}=2$; $N_{10'}=22$),  Minni341 ($N_{3'}=2$; $N_{10'}=11$),  Minni342 ($N_{3'}=1$; $N_{10'}=5$),  Minni343 ($N_{3'}=3$; $N_{10'}=13$) and Minni344 ($N_{3'}=1$; $N_{10'}=15$). \\
Even though this may appear to be an inconsistency,  there is varied observational evidence shown the presence of RR Lyrae stars also in some metal-rich GCs,  such as NGC~6388 ($[Fe/H] = -0.44 $), NGC~6441 ($[Fe/H] = -0.46$ -- \citealt{Pritzl2002,Clementini2005}),  NGC~6440 ($[Fe/H] = -0.36$), and Patchick~99 ($[Fe/H] = -0.20$ --  \citealt{Garro2021a}).\\

We also computed the specific frequency of RR Lyrae stars in the Sgr GCs.  \cite{Suntzeff1991}  defined the specific frequency of RR Lyrae stars $S_{RR}$ as the number of RR Lyrae stars $N_{RR}$ per unit luminosity, normalised to a typical Galactic globular cluster luminosity of $M_{V_t}=-7.5$ mag:

\begin{equation}
S_{RR} = N_{RR} / 10^{-0.4(M_{V_t} + 7.5)}
\end{equation}
(following the notation of \cite{Harris1996}).  Using the $V-$band total luminosities and considering the RR Lyrae stars within $3$ arcmin,  we find $4 \lesssim S_{RR} \lesssim 87$ for the Sgr GCs.
Also, they are faint GCs and it is therefore not surprising that there are low number statistics. Indeed, we note that $S_{RR}$ are lower limits,  because we have not included information about detection completeness, or counted the candidate variable stars.  On the other hand,  in calculating $S_{RR}$ we have normalized the $N_{RR}$ values to full cluster luminosities.  Accounting for this effect is not trivial, because we do not know the spatial distribution of RR Lyrae stars in any clusters and  given that we have imaged the centre of each cluster, we expect the $N_{RR}$ to be 80--90 per cent complete. Hence, the $S_{RR}$ values may be 10--20 per cent greater than quoted above. Therefore, the determination of these quite reliable $S_{RR}$ allow us to give a lower age limit of $Age \sim 10$ Gyr for all clusters with $S_{RR}>0$.  However, the absence of RR Lyrae stars does not necessary imply a young ages,  because RR Lyrae are uncommon in metal-rich GCs for example. \\
At this point, we make comparison with the MW from the 2010 compilation of \cite{Harris1996} catalogue.   That catalogue contains only four of the $\sim 150$ Galactic GCs as having $S_{RR} > 60$, and only two of these have $S_{RR} > 100$. The largest value, that for Palomar 13, is $S_{RR} = 127.5$.   According to this and given our sample incompleteness,  it is very likely that Sgr dwarf galaxy has $S_{RR}$ greater than this.  It is certainly intriguing that only a tiny fraction of Galactic GCs have very high $S_{RR}$ as well as these Sgr GCs since we find only one GC with $S_{RR}> 60$ (Minni145),  and any with $S_{RR} > 100$.  This may suggest that Sgr GCs follow a similar trend as of MW GCs.

\section{Discussion}
In the next sections, we highlight the main differences between the GCs within the Sgr dSph itself, comparing the resulting values for the newly discovered GCs (Table \ref{tablegcs}) and the well-known Sgr GCs (Table \ref{tablecomp}).  However, we want also to broaden the discussion by making the first comparison between the Sgr GC system and the GC system of neighbouring galaxies: the MW, M31 and LMC.

\subsection{The Sagittarius globular cluster system}
From \cite{Bellazzini2020} at least nine GCs are associated to the Sgr stream: four clusters in the Sgr remnant – NGC 6715, Terzan 7, Terzan 8, Arp 2; two clusters in the trailing arm – Palomar 12 and Whiting 1; and three clusters likely associated to an old arm - NGC 2419, NGC~1447 and NGC~5634.   Actually, also Berkeley 29 an Saurer 1, two younger clusters, originally associated with the Sgr tidal extension, have now been discarded using the improved \textit{Gaia} EDR3 PMs \citep{GaiaCollaboration2020}. M21 increased the number of GCs in the Sgr system. Hence,  these discoveries pave the way to multiple studies on the understanding of Sgr dwarf itself, on chemistry and dynamics of these systems, especially when compared with other GC systems,  and they can also help to guide theoretical studies and simulations (e.g.,\citealt{Vasiliev2021a,Vasiliev2021b}). \\

\begin{figure}
\centering
\includegraphics[width=8cm, height=10cm]{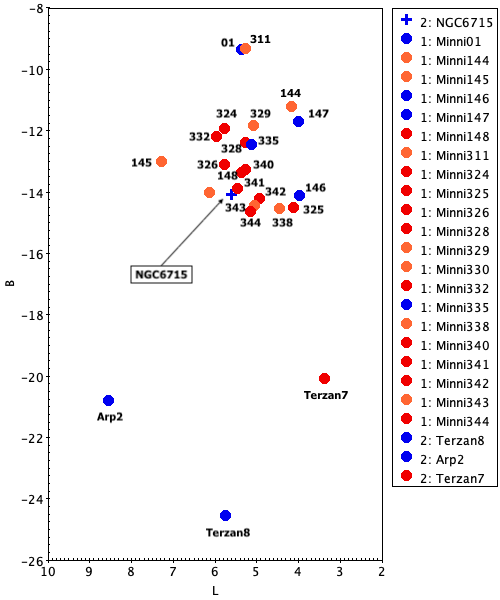}
\caption{Latitude versus longitude map depicting the position of Sgr GCs in the main body.   The coloured points highlight the cluster metallicity,  adopting red for GCs with $[Fe/H]\geqslant -0.5$,  orange for GCs with $-1.0\leqslant[Fe/H]\leqslant -0.6$,  and blue for GCs with $[Fe/H]\leqslant -1.1$ for our convention. We included NGC~6715, Terzan~7, Terzan~8 and Arp~2 for comparison. }
\label{position}
\end{figure}

\begin{figure}
\centering
\includegraphics[width=9cm, height=7cm]{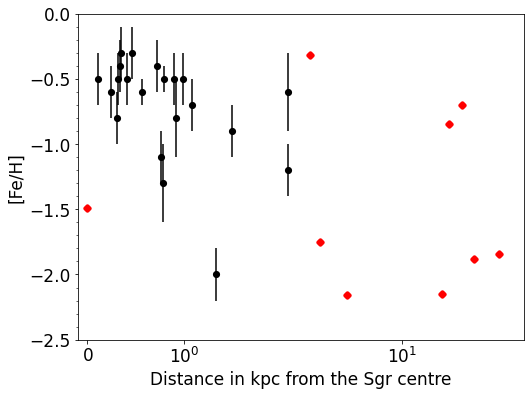}
\caption{Radial dependence of the metallicity for Sgr GCs.  We consider the projected distance from the centre of Sgr, adopting $D = 26.5$ kpc for this galaxy.  The new discovered Sgr GCs are marked with black points, while the known Sgr GCs are represented as red diamonds.}
\label{radialmetallicity}
\end{figure}

\begin{figure}
\centering
\includegraphics[width=9cm, height=7cm]{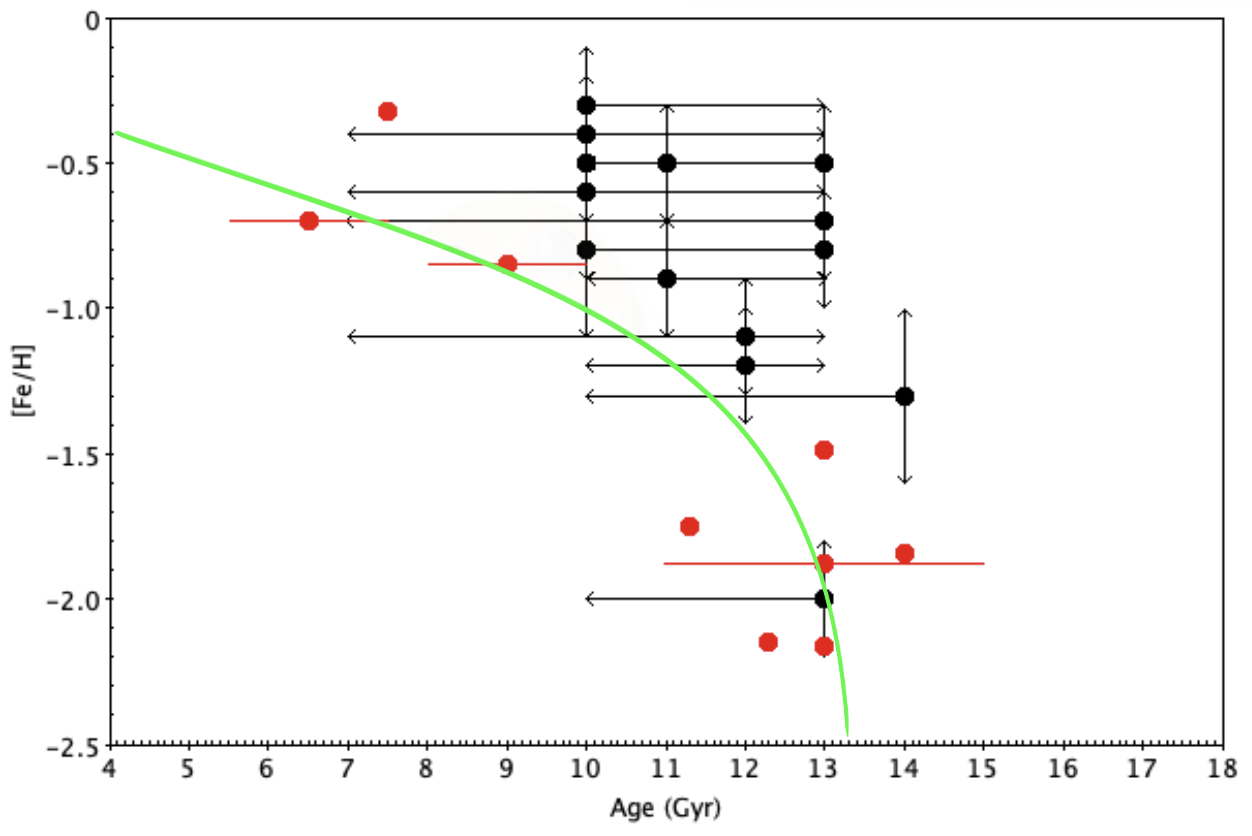}
\includegraphics[width=9cm, height=7cm]{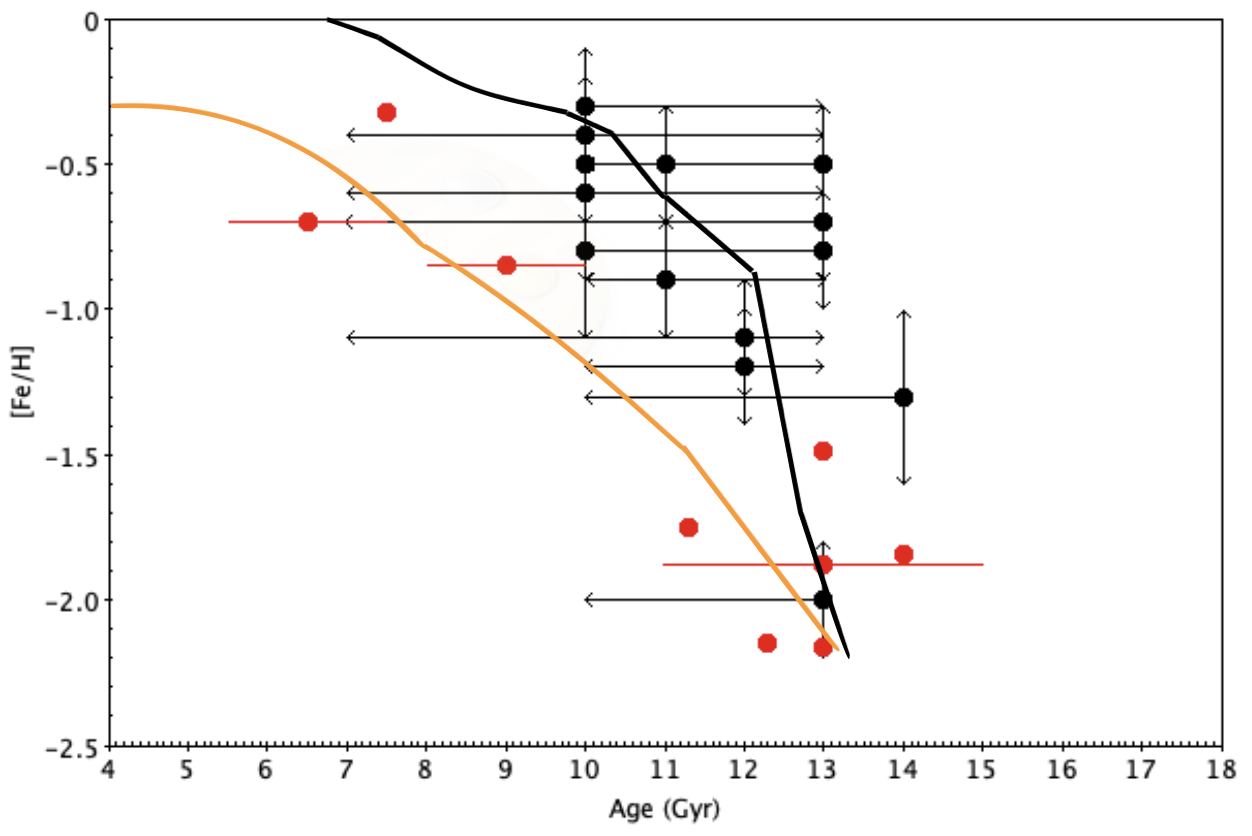}
\caption{Age-metallicity relation for the Sgr GC system.  The newly discovered Sgr GCs are marked with black points, while the known Sgr GCs are represented as red diamonds.  \textit{On the top panel:} the green line reproduces the AMR for the MW halo GCs shown in \cite{Forbes2020} (see their Fig.  3).  \textit{On the bottom panel:} the solid lines reproduce the median star cluster AMR relation for LMC (orange) and Milky Way-mass (black) galaxies in the E-MOSAICS simulations by \cite{Horta2021} (see their Fig.  3). }
\label{age_metall}
\end{figure}

We find that the main body of the Sgr galaxy is mainly a metal-rich component as shown in Fig.  \ref{position}.  Figure \ref{radialmetallicity} shows the relation between the distance to the Sgr center and the metallicity of its GCs ranging from NGC 6715 (which coincides with the Sgr nucleus) to the most distant GC NGC~2419.  We note that, apart from NGC 6715, which is metal-poor,  the metallicity is $-0.3>[Fe/H]>-0.9$ in the innermost regions ($R<0.6$ kpc), whereas between $R\approx 0.6$ kpc and $40$ kpc we can see that there are both metal-rich (with the same metallicity range as the inner part) and metal-poor GCs, with $-1.1>[Fe/H]>-2.3$.  The spread in the MD of Sgr GCs seems to suggest that the Sgr dSph had an extended star formation history and therefore now contains high metallicity clusters \citep{Hughes2019}.  
It appears that a metallicity gradient could be hinted at since we do not find metal-poor GCs in the innermost regions, but we cannot conclude with extreme certainty that this occurs in the main body of Sgr dSph.
Probably the star cluster formation in the Sgr galaxy occurred in two different episodes. From the AMR diagram (Fig. \ref{age_metall}), we see that metal-poor GCs represent the old component with an average age $ \sim 13 $ Gyr, while the metal-rich GCs span a wider range of ages from younger with $ t \sim 7-8 $ Gyr to older with $ t\sim 10-14 $ Gyr.   This result also qualitatively agree with the Sgr star formation history by Hasselquist et al.  2021 in preparation,  as they demonstrated that Sgr formed its metal-rich ($-0.9 <$ [Fe/H] $< -0.3$) GCs some 6-8 Gyr ago.  However, we can only speculate on the probable formation of these clusters, since the epoch of accretion of the Sgr dwarf is largely uncertain and our GC ages are approximative: the older objects could be formed \textit{in-situ} within the Sgr progenitor and have been consequently accreted onto the MW, whereas the younger ones can be the result of the merging event with the MW.  This could be distorted since when a satellite galaxy infalls into the main galaxy halo, a subsequent gas stripping leads to a truncation of the GC formation in the smaller galaxy. However, there is evidence that some galaxies (i.e., LMC and SMC) continue to form clusters after they have entered the halo of the main galaxy.  Additionally,  \cite{Hughes2019}, using the E-MOSAIC simulation, found that more massive galaxies can continue to form GCs for longer after entering the halo of the main galaxy.  Many of the satellite galaxies in this population produce streams because the galaxies were accreted later and so the streams survive until present day.  More massive streams host younger and more-metal rich GCs.\\
Although our age estimates are rough and we need deeper observations in order to reach the MSTO,  Fig.  \ref{age_metall} can help us to broadly reconstruct the formation history of the Sgr galaxy.  Indeed, we find similarities with the AMR shown in \cite{Massari2019} and \cite{Forbes2020}.  Both these works assumed a leaky-box age-metallicity relation\footnote{The form of the age-metallicity relation used by \cite{Forbes2020} is $[Fe/H]=-p\ \ln (t/t_f)$ where $p$ is the effective yield of the system and $t_f$ is the look-back time when the system first formed from non-enriched gas.}.  We have reproduced in Fig.  \ref{age_metall} the best-fit halo AMR found by \cite{Forbes2020}.  We notice that Forbes’ fit follows the red diamonds, which represent the known Sgr GCs, whereas the new discovered ones are located above that best-fit.  On the other hand, we also compared the Sgr AMR with those of the MW and LMC, as shown in Fig.  \ref{age_metall}. We appreciate that the Sgr AMR looks more like the MW  \citep{Leaman2013,Horta2021}, whereas the halo AMR by \cite{Forbes2020} shows a similar trend of the LMC AMR by \cite{Horta2021}.  It is important to note that when \cite{Horta2021} compared the AMR of massive clusters for the LMC sample, they argued that the AMR for the satellite galaxies is systematically above that for the MW as expected from mass assembly bias. Investigating on the origin of this bias, they found that the origin of the AMR trend in the stellar populations is a function of the galaxy mass. This means that higher mass galaxies undergo more rapid enrichments, as they retain the SNe ejecta and consequently this leads to a rapid enrichment of the interstellar medium (ISM). Instead, lower mass galaxies grow more slowly over time, as their potential wells are not deep enough and large fraction of the SNe ejecta and stellar mass are lost, leading to a slower metal enrichment process. Therefore, this seems to indicate that the Sgr progenitor was not a small galaxy, on the contrary it was massive enough to retain the stellar mass loss to enrich its ISM quickly.

 \begin{figure}
\centering
\includegraphics[width=8cm, height=6cm]{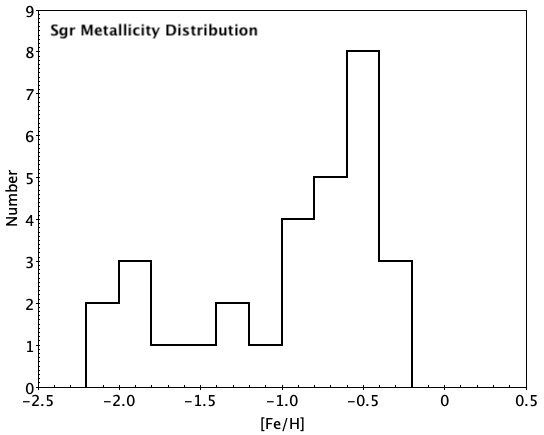}
\includegraphics[width=8cm, height=6cm]{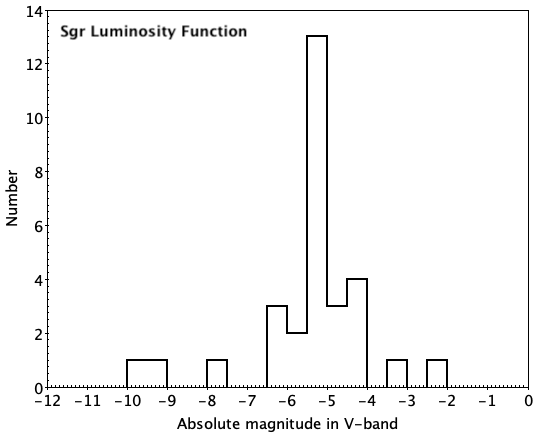}
\caption{Sagittarius metallicity distribution (top panel) and luminosity function (bottom panel), including both new and known GCs.}
\label{Sgr_function}
\end{figure}

\subsection{Metallicity distribution}
The metallicity distribution (MD) can provide important clues as to the process and conditions relevant to galaxy formation. Therefore, comparison with other systems can lay out a wider view. \\ 
For these purposes, we build-up the MD for the Sgr GC system as shown in Fig.  \ref{Sgr_function}. We notice two separate peaks, one metal-rich at $[Fe/H]_{MR}^{Sgr}=-0.56\pm 0.18$~dex and the second metal-poor at $[Fe/H]_{MP}^{Sgr}=-1.75 \pm 0.35$~dex.  Additionally,  we construct the MD for the MW GC system based on the 2010 compilation of the McMaster Milky Way Globular Cluster catalogue \cite{Harris1996},  including 157 GCs.  We also involve the M31 GC system,  including 504 GCs by \cite{Fan2008}.  Finally,  we also considered the LMC satellite galaxy,  but in this case we did not find a complete catalogue,  so we preferred to make it ourselves by referring to \cite{Mackey2004, McLaughlin2005, Lyubenova2010, Colucci2011, Noel_2013, Jeon_2014,WagnerKaiser2017,Piatti2018,  Piatti2019, Horta2021}. We selected 41 LMC GCs, listed in Table \ref{LMC}, where we summarized: cluster ID, observed $V$ magnitude, reddening, metallicity, age,  absolute magnitude in $V-$band and references. Although many additional star clusters have been catalogued by \cite{Palma2016}, we did not consider them as most of these objects show very young age and a mean metallicity value of $\sim -0.3$ dex,  and also because they have not been clearly categorized as open or globular clusters. \\

\begin{table*}
\centering 
 \begin{adjustbox}{max width=\textwidth}
\begin{tabular}{lcccccl}
\hline\hline
Cluster ID         &  Vmag  & $E(B-V)$     &  [Fe/H]    &  Age   & $M_{Vtot}$ &   References \\
        & [mag] & [mag] &  [dex] & [Gyr] & [mag] & \\
\hline
NGC 1466    &   $ 11.59 $  &$0.09 $  &$  -1.70   $ &$ 11.00  $&$  -7.25 $      &\cite{WagnerKaiser2017,Jeon_2014} (WK17, J14) \\
NGC 1711    &   $ 10.11 $  &$0.09 $  &$  -0.57   $ &$ 0.05   $&$  -8.73 $      &\cite{Colucci2011} (C11)\\
NGC 1754    &   $ 11.57 $  &$0.07 $  &$  -1.50   $ &$ 12.96  $&$  -7.21 $      &\cite{Piatti2018, Piatti2019,Lyubenova2010} (P18, P19, L10)\\
NGC 1786    &   $ 9.50  $  &$ 0.07$  &$   -1.75  $ &$  12.30 $&$  -9.28 $      &P18, P19\\
NGC 1718    &   $ 12.25 $  &$0.10 $  &$   -0.40  $ &$   2.0  $&$  -6.62 $      & \cite{Horta2021, Kerber2007} (H21,K07)\\ 
NGC 1835    &   $ 10.60 $  &$ 0.14$  &$   -1.72  $ &$  13.37 $&$  -8.40 $      &P18, P19\\
NGC 1841    &   $ 11.43 $  &$0.14 $  &$  -2.02   $ &$ 12.57  $&$  -7.57 $      &P18, P19, H21, \cite{Grcevich2009} (G09)\\
NGC 1866    &   $ 9.89  $  &$0.06 $  &$  -0.51   $ &$ 0.13   $&$  -8.86 $      &C11\\
NGC 1898    &   $ 11.86 $  &$0.07 $  &$  -1.32   $ &$ 13.50  $&$  -6.92 $      &P18, P19\\
NGC 1916    &   $ 10.38 $  &$0.13 $  &$  -1.54   $ &$ 12.56  $&$  -8.58 $      &P19, C11\\
NGC 1928    &   $ 12.47 $  &$0.08   $  &$  -1.30   $ &$ 13.50  $&$  -6.34    $    &P19, \cite{Mackey2004}\\
NGC 1939    &   $ 11.83 $  &$0.16  $  &$  -2.00   $ &$ 13.50  $&$  -7.23    $    &P19, \cite{Mackey2004}\\
NGC 1978    &   $ 10.74 $  &$0.09 $  &$  -0.38   $ &$ 1.9    $&$  -8.10 $      &C11\\
NGC 2002    &   $ 10.10 $  &$0.20 $  &$  -2.20   $ &$  0.18  $&$  -9.08 $      &C11\\
NGC 2005    &   $ 11.57 $  &$0.07 $  &$  -1.74   $ &$ 13.77  $&$  -7.21 $      &P18, P19, C11, L09\\
NGC 2019    &   $ 10.95 $  &$0.07 $  &$  -1.56   $ &$ 16.20  $&$  -7.83 $      &P18, P19, C11\\
NGC 2100    &   $ 9.60  $  &$0.26 $  &$  -0.32   $ &$ 0.015  $&$  -9.77 $      &C11\\
NGC 2210    &   $ 10.94 $  &$0.10 $  &$  -1.55   $ &$ 10.43  $&$  -7.93 $      &P18, P19\\
NGC 2257    &   $ 12.62 $  &$0.05 $  &$  -1.77   $ &$ 11.54  $&$  -6.09 $      &P18, P19, G06, WK17\\
ESO 121-SC3 &   $ 14.04 $  &$0.04 $  &$  -1.05   $ &$ 8.50   $&$  -4.64 $      & P18, P19\\
Reticulum   &   $ 14.25 $  &$0.03 $  &$  -1.47   $ &$ 11.9   $&$  -4.40 $      & G06, J14, WK17\\
Hodge11     &   $ 11.93 $  &$0.08 $  &$  -2.00   $ &$ 13.92  $&$  -6.88 $      &G06,WK17, P18, P19 \\
NGC1651     &   $ 12.43    $  &$0.13 $  &$  -0.70   $ &$   2.0  $&$  -6.53    $    & H21, K07, \cite{Noel_2013} (N13)\\
NGC 1777    &   $ 12.41    $  &$0.13 $  &$  -0.60   $ &$   1.1  $&$  -6.55    $    & H21, K07, N13\\
NGC 1783    &   $ 10.60    $  &$0.10   $  &$  -0.35   $ &$   1.7  $&$  -8.26    $    & H21,\cite{Mucciarelli2008}, N13\\
NGC 1806    &   $  11.12    $  &$0.08   $  &$  -0.60   $ &$   1.5  $&$  -7.69    $    & H21, \cite{Mucciarelli2014}, N13\\
NGC 1831    &   $ 10.70    $  &$0.13 $  &$  -0.10   $ &$   0.7  $&$  -8.26    $    & H21, K07, N13\\
NGC 1856    &   $ 9.85    $  &$0.07 $  &$  -0.40   $ &$   0.3  $&$  -8.93    $    & H21, K07, N13\\
NGC 1868    &   $ 11.14   $  &$0.13 $  &$  -0.70   $ &$   1.1  $&$  -7.82    $    & H21, K07, N13 \\
NGC 2121    &   $ 11.84    $  &$0.17 $  &$  -0.40   $ &$   2.9  $&$  -7.25    $    & H21, K07, N13\\
NGC 2136    &   $ 10.20    $  &$0.10  $  &$  -0.40   $ &$   0.09 $&$  -8.67    $    & H21, N13 \cite{Mucciarelli2012}\\
NGC 2137    &   $ 11.94    $  &$0.03   $  &$  -0.40   $ &$   0.09 $&$  -6.71     $    &  H21, \cite{Mucciarelli2012,McLaughlin2005}\\
NGC 2155    &   $ 12.27    $  &$0.14$  &$  -0.35   $ &$   2.5  $&$  -6.73    $    & H21, N13,  \cite{Martocchia2019}, \\
NGC 2162    &   $ 12.22    $  &$0.13 $  &$  -0.40   $ &$   1.2  $&$  -6.74    $    & H21, K07, N13\\
NGC 2173    &   $ 11.92    $  &$0.13 $  &$  -0.60   $ &$   1.6  $&$  -7.04    $    & H21, K07, N13\\
NGC 2209    &   $ 12.76    $  &$0.13 $  &$  -0.50   $ &$   1.2  $&$  -6.20    $    & H21, K07, N13\\
NGC 2213    &   $  12.08    $  &$0.13 $  &$  -0.70   $ &$   1.7  $&$  -6.88    $    & H21, K07, N13\\
NGC 2249    &   $  11.84    $  &$0.13 $  &$  -0.40   $ &$   1.0  $&$   -7.12    $    & H21, K07, N13\\
Hodge 6     &   $ --    $  &$--   $  &$  -0.35   $ &$   2.0  $&$  --    $    &  H21,  \cite{Hollyhead2019}\\
SL 506      &   $ --    $  &$0.08 $  &$  -0.40   $ &$   2.2  $&$  --    $    & H21, K07\\
SL 663      &   $ --    $  &$0.07 $  &$  -0.70   $ &$   3.1  $&$  --    $    & H21, K07\\ 
\hline\hline
\end{tabular}
\end{adjustbox}
\caption{Properties of the LMC GCs used for comparison.}
\label{LMC}
\end{table*}

\begin{figure*}
\centering
\includegraphics[width=6cm, height=4.5cm]{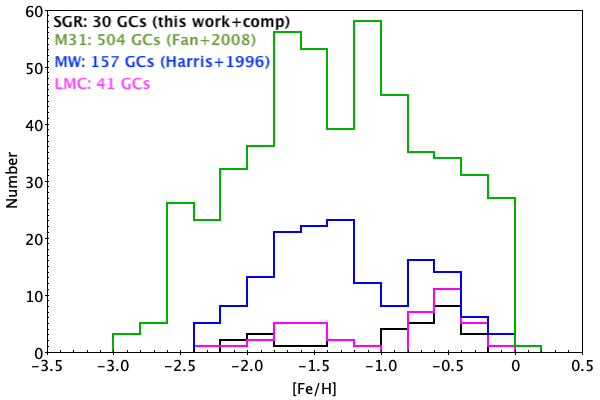}
\includegraphics[width=6cm, height=4.5cm]{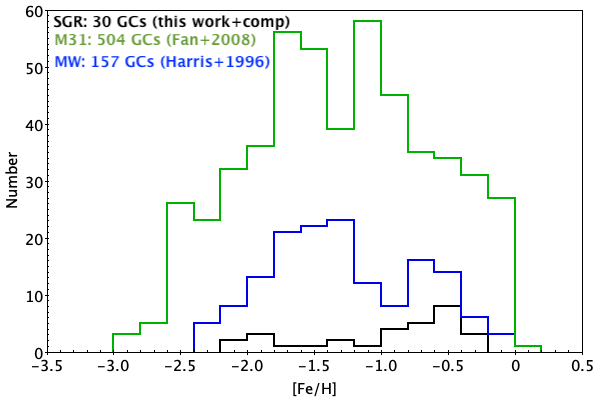}
\includegraphics[width=6cm, height=4.5cm]{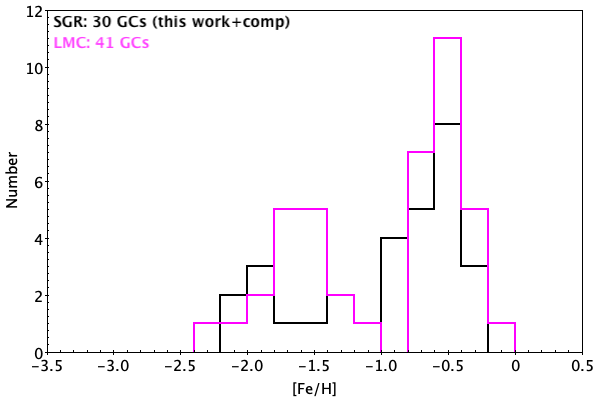}
\caption{Metallicity distribution for Sgr (black histogram),  MW (blue histogram),  M31 (green histogram) and LMC (magenta histogram) GC systems (in the left panel).  We also show separately: the Sgr and LMC metallicity distributions (in the right panel) as a zoomed figure of the left panel; while we use the middle panel in order to emphasize the Sgr MD.  The histograms are constructed adopting a bin size of 0.2 dex.}
\label{mf}
\end{figure*}

Figure \ref{mf} shows the MD for each sample: Sgr, MW, M31 and LMC.  We find that each MD shows a Gaussian-like distribution.  Especially,  a bimodal MD is shown for the Andromeda and MW galaxies,  with the metal-poor peaks at $[Fe/H]_{MP}^{M31}~ =~ -1.71 \pm 0.46 $ dex and $[Fe/H]_{MP}^{MW}~=~ -1.55 \pm 0.35$ dex and metal-rich peaks at $[Fe/H]_{MR}^{M31} = -0.76 \pm 0.39$ dex and $[Fe/H]_{MR}^{MW} = -0.54 \pm 0.22$ dex, respectively.  Whereas we find a double-peaked distribution for LMC with $[Fe/H]_{MP}^{LMC}~ =~ -1.66 \pm 0.30 $ dex and $[Fe/H]_{MR}^{LMC}~ =~ -0.47 \pm 0.15 $~dex.  We expected a bimodal MD, as it is a typical feature of all GC systems (e.g. \citealt{Ashman1992}).\\ 

From the comparisons between these galaxies, many differences can be noted.  First,  the total number of metal-poor Sgr GCs is smaller than that of the LMC, while it is similar when only metal-rich populations are considered.  
On the other hand,  the Sgr MD seems to follow both the MW and M31 MDs.  In particular, the Sgr metal-rich GCs look more like the MW bulge GCs, showing also similar wide age range.  Therefore at odd with the LMC, all the metal-rich GCs in the Sgr dwarf are relatively old (Age $>6$ Gyr). This is consistent with the hypothesis that the Sgr GCs formation occurred after infalling into the MW halo and thus the progenitor satellite was gas-rich, in accordance with what is shown in Fig.  \ref{age_metall}.  \cite{Kruijssen2011}, based on numerical simulations of isolated and merging disc galaxies, found that the star clusters that survive the merger and populate the merger remnants are typically formed at the moments of the pericentre passage,  namely slightly before the starbursts that occur during a galaxy merger.  These clusters constitute a large fraction ($30-60$ per cent per pericentre passage) of the survivors. They survived for two reasons: firstly,  they are formed before the peak of the starbust,  and secondly the formed clusters, during the pericentre passage, are ejected into the stellar halo, where the disruption rate is low and the survival chance is high. However, the clusters that are produced in the central regions during the peak of the starburst are short-lived and disrupt before they can migrate to the halo.   

\subsection{Luminosity function}
The brightness distribution of the GCs, known as the globular cluster luminosity function (GCLF), is an important tool, which can be used, as a distance indicator as well as to constrain the theories on the formation and evolution of GCs (e.g, \citealt{DeGrijs2005, Nantais_2006, Rejkuba2012}), and more precisely to predict the dynamical processes that come into play in the destruction of GCs, especially when merging events between galaxies are involved.  The main dynamical processes that can destroy GCs are: \textit{(i)} the mass loss due to the stellar evolution (i.e. , supernovae explosions --  \citealt{Lamers2010}); \textit{(ii)} dynamical friction \citep{Alessandrini2014}; \textit{(iii)} tidal shocks heating by passages through bulge or disk \citep{Gnedin1997,Piatti_Carballo2020};  and \textit{(iv)} evaporation due to 2-body relaxation \citep{Madrid2017}.  All of these processes, especially the last two, can change the luminosity distribution. \\ 
Many works (\citealt{Rejkuba2012}, and references therein) have demonstrated that the GCLF is not a Gaussian distribution,  nor there is a "Universal distribution" in most galaxies \citep{Huxor2014}.  Although the Gaussian fit is a useful parametrization (especially at the brighter end), its use has no physical motivation. Indeed,  the shape of the GCLF deviates from the Gaussian symmetric distribution for the MW as well as the external galaxies, since it shows a longer tail towards the faint end.  This asymmetry is strongly related to the dynamical evolution of the initial cluster luminosity function, which is well-approximated by power laws of the form $dN(L)/dL \propto L^{\alpha}$, with slopes in the range $-2.4 \lesssim \alpha \lesssim -2.0$ as found by i.e.  \cite{Larsen_2002}. Another reason that could explain why the GCLF deviates from the symmetry can also be found in the mass function, since mass and luminosity are related parameters in GCs. So it is expected that as dynamic evolution modifies the GCLF, it also has an effect on the mass function (e.g. \citealt{Fall_2001}).
We suggest that many "inconclusive" objects,  defined by M21 because they require additional data for confirmation or rejection (Minni02, 03, 04, 312, 313, just to name a few), could be widespread or dissolute GCs.\\

\begin{figure*}
\centering
\includegraphics[width=6cm, height=4.5cm]{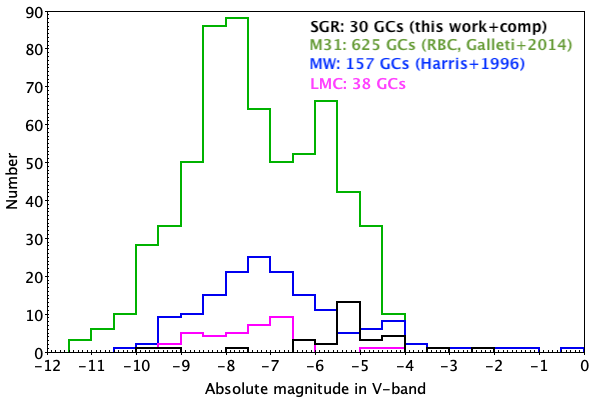}
\includegraphics[width=6cm, height=4.5cm]{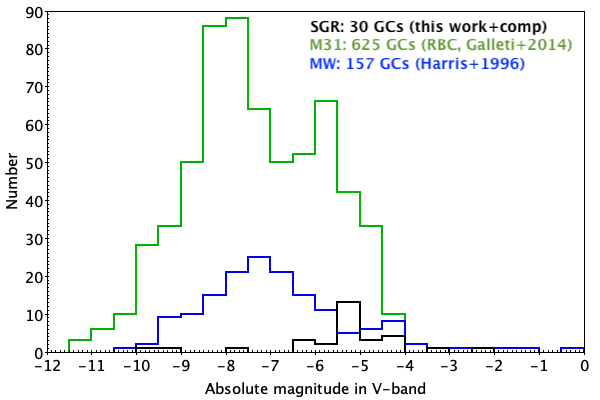}
\includegraphics[width=6cm, height=4.5cm]{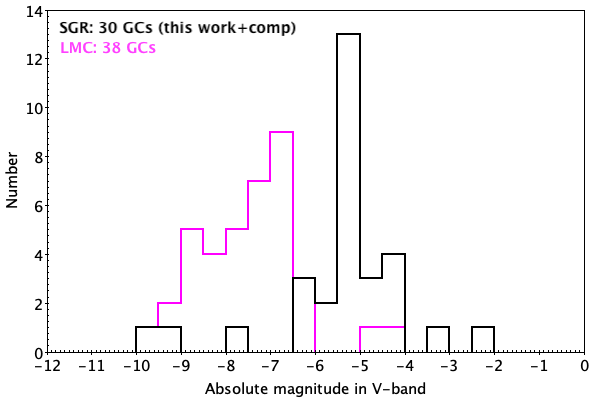}
\caption{Luminosity function for Sgr (black histogram),  MW (blue histogram),  M31 (green histogram) and LMC (magenta histogram) GC systems (in the left panel).  We also show separately: the Sgr and LMC luminosity functions (in the right panel) as a zoomed figure of the left panel; while we use the middle panel in order to emphasize the Sgr LF. The histograms are constructed considering a bin size of 0.5 mag.}
\label{lf}
\end{figure*}

Following the same line as for the MD, we compare the Sgr GCLF with the MW, M31 and LMC GCLFs.  \\
Briefly, we describe the method used to obtain reliable absolute magnitudes for the GCs in each sample. We derive the absolute magnitude in $V-$band for our Sgr GCs as explained in Section 3 and shown in Table \ref{tablegcs}, and we use the luminosities for the known Sgr GCs, shown in Table \ref{tablecomp}.  Figure \ref{Sgr_function} depicts the final Sgr GCLF.  This is a unimodal distribution, which exhibits a prominent peak at $M_V^{Sgr}~=~-5.46\pm 1.46$ mag, and shows a tail toward the brighter end. \\
To build the MW GCLF, we adopt the $M_{V}$ values from the 2010 compilation of the \cite{Harris1996} catalogue.  For the LMC GCLF,  we use again our LMC catalogue with 41 GCs (Table \ref{LMC}),  for which we have also recovered, besides metallicities,  their $V$ magnitudes and their colour excess $E(B-V)$.  We adopt the LMC distance modulus $(m-M)_{0}=18.56$ mag from \cite{DiBenedetto2008}.  For the M31 GCLF, we prefer a more complete catalogue: the Revised Bologna Catalogue (RBC, V.5, August 2012 -- \citealt{Galleti2006,Galleti2014}) of M31 globular clusters and candidates.  This catalogue includes 2,060 objects, but we selected the 625 confirmed GCs ($flag=1$), with the optical photometry.  For M31 we use a distance modulus $(m-M)_{0}~=~24.47$~mag \citep{McConnachie2005}.  At this point, we are able to calculate the absolute magnitude in $V-$band for LMC and M31 GC system, applying the same method as for Sgr GCs (Section 3).\\
Once obtained the GC luminosities, we can build up the luminosity function for each sample, thus making comparisons between these galaxies. \\
In Fig. \ref{lf},  we can clearly see a double-peaked M31 GCLF with the fainter peak at $M_{V}^{M31}=-5.71\pm 0.71$ mag and the brighter peak at $M_{V}^{M31}=-7.95\pm 1.12$ mag.  A less pronounced double-peaked is notable for the MW GCLF, which is shifted by $\sim 0.5$ mag toward fainter luminosities with respect to that of M31,  since the fainter peak is at $M_{V}^{MW}=-4.01\pm 1.28$ mag and the brighter one is at $M_{V}^{MW}=-7.46\pm 1.04$ mag.  On the other hand, we find that the LMC GCLF is a unimodal distribution peaked at $M_{V}^{LMC}=-7.43\pm 1.17$ mag,  based on a sample with 38 GCs for which we recovered both $V$ magnitudes and reddening values. \\
A plausible reason to explain the observed double-peaked GCLFs in the MW and M31 could be that many GCs have been accreted by the Andromeda halo as well as by the MW halo \citep{Peacock2010, Huxor2014}.  
This scenario finds an additional support when we include the Sgr GC system, where the LF shows a fainter peak.  More generally,  \cite{Mackey2005} found a similar fainter peak in the GCLF of the "young halo" GCs of the MW, which they argued are most likely accreted objects.  Therefore, this seems to hint that accreted Sgr GCs have survived disruption processes deriving from the merging event, and those GCs that  we see today are the remnants of more massive objects or the final products of the accretion. However,  we note that the faint Sgr clusters are relatively more numerous than their MW counterparts. Therefore this can indicate dynamical evolution, such as more efficient destruction in the larger galaxies, or incompleteness at the fainter peak,  since we did not include many of the low luminosity Galactic GCs recently discovered,  nor we considered that many more faint objects are still to be found. \\
Finally,  the actual LMC LF may be more complex than the one we show in Fig.  \ref{lf},  since there is evidence for a very complex and still ongoing star formation activity \citep{Bruzual2003}. However based on our data, we can see that the LMC $M_V$ GCLF peak is brighter than the Sgr GCLF  peak.  This seems to suggest that LMC forms more massive clusters or that the dynamical processes in the LMC are different than those undergone by the Sgr dwarf. However, this could also be due to our results suffering from incompleteness, or to the overall LMC youth when compared with the Sgr dSph.
\section{Summary and conclusions}
We analyse the new 21 GCs recently discovered by M21 in the main body of Sgr dwarf galaxy. We built their optical \textit{Gaia} EDR3 and near-IR VVVX CMDs and proceed to fit isochrone models in order to derive suitable values for their metallicities and ages.  We confirm our metallicity estimates using an independent method ($slope_{RGB} - [Fe/H]$ relation by \citealt{Cohen_2015}). Furthermore,  we also calculated for each target their total luminosity in the $K_s-$ and $V-$bands,  finding all of them to be faint ($M_V > -6.2$) clusters.  \\

Once metallicities and luminosities were obtained, we were able to build up the MD and the LF for the Sgr GC system (including the new GCs discovered by M21 and the previously known GCs associated to the Sgr dwarf) for the first time.  We find that the main body of Sgr contains prevalently a metal-rich component. However, both metal-poor and metal-rich GCs are found, with very different ages: metal-poor GCs are old with $t\sim 13$ Gyr, whereas  metal-rich GCs show a wider age range from $7-8$ Gyr up to $10-13$ Gyr. Our age estimates are not very accurate, since the MSTO for the GCs in our sample is below our detection limit,  but we derived minimum ages assigning an Age $>10$ Gyr for those GCs with $S_{RR}>0$, and also a lower limit of $\sim 7$ Gyr (derived from $\Delta K_{s}$(HB-MSTO) for all GCs with $S_{RR}=0$.  We did not detect a metallicity gradient, but we find that the innermost regions ($R<0.6$ kpc) show GCs with metallicities of $\sim -0.5$ dex, while between 0.6 and 40 kpc from the Sgr nucleus we can find both metal-rich and metal-poor populations.\\
We have done a comparison with the MD and GCLF of other galaxies: the MW, Andromeda galaxy and LMC, thus placing constrains on the formation and evolution of the Sgr dwarf galaxy.  Based on this comparison,  we suggest that the Sgr  progenitor could have been a gas-rich galaxy,  and this gas was retained and subsequently converted into GCs during the infall into the MW halo.  Many GCs survived the main dynamical processes (i.e. tidal shocks and two-body relaxation process), probably because they were formed before the main star formation burst and pulled toward the halo where these processes are less efficient.  If this mechanism occurred, we may believe that some of the "inconclusive" objects, detected in the main body of Sgr by M21, could be dissolved GCs associated to the Sgr dwarf.  Additionally, when we compare the GCLFs, since the Sgr distribution peaks at lower luminosities ($M_V\approx -5.5$ mag) than all other samples, we can conclude that the dynamical processes that destroy GCs are more efficient in larger galaxies than in smaller ones, or that many faint GCs are missed in the present compilations.

\newpage
\begin{acknowledgements}
We gratefully acknowledge the use of data from the ESO Public Survey program IDs 179.B-2002 and 198.B-2004 taken with the VISTA telescope and data products from the Cambridge Astronomical Survey Unit.  ERG acknowledges support from an UNAB PhD scholarship and ANID PhD scholarship No. 21210330.  D.M. acknowledges support by the BASAL Center for Astrophysics and Associated Technologies (CATA) through grant AFB 170002.  J.A.-G. acknowledges support from Fondecyt Regular 1201490 and from ANID -- Millennium Science Initiative Program -- ICN12\_009 awarded to the Millennium Institute of Astrophysics MAS.
\end{acknowledgements}

\bibliographystyle{aa.bst}
\bibliography{bibliopaper}

\begin{thebibliography}{139}
\expandafter\ifx\csname natexlab\endcsname\relax\def\natexlab#1{#1}\fi

\bibitem[{Abraham {et~al.}(2018)Abraham, Danieli, van Dokkum, Conroy,
  Kruijssen, Cohen, Merritt, Zhang, Lokhorst, Mowla, Brodie, Romanowsky, \&
  Janssens}]{Abraham_2018}
Abraham, R., Danieli, S., van Dokkum, P., {et~al.} 2018, Research Notes of the
  {AAS}, 2, 16

\bibitem[{{Alessandrini} {et~al.}(2014){Alessandrini}, {Lanzoni}, {Miocchi},
  {Ciotti}, \& {Ferraro}}]{Alessandrini2014}
{Alessandrini}, E., {Lanzoni}, B., {Miocchi}, P., {Ciotti}, L., \& {Ferraro},
  F.~R. 2014, \apj, 795, 169

\bibitem[{{Alonso-Garc{\'\i}a} {et~al.}(2018){Alonso-Garc{\'\i}a}, {Saito},
  {Hempel}, {Minniti}, {Pullen}, {Catelan}, {Ramos}, {Cross}, {Gonzalez},
  {Lucas}, {Palma}, {Valenti}, \& {Zoccali}}]{AlonsoGarcia2018}
{Alonso-Garc{\'\i}a}, J., {Saito}, R.~K., {Hempel}, M., {et~al.} 2018, \aap,
  619, A4

\bibitem[{{Ashman} \& {Zepf}(1992)}]{Ashman1992}
{Ashman}, K.~M. \& {Zepf}, S.~E. 1992, \apj, 384, 50

\bibitem[{Ashman \& Zepf(1998)}]{ashman_1998}
Ashman, K.~M. \& Zepf, S.~E. 1998, Cambridge Univ. Press, Cambridge (Cambridge
  Astrophysics Series; 30)

\bibitem[{{Barbuy} {et~al.}(2006){Barbuy}, {Bica}, {Ortolani}, \&
  {Bonatto}}]{Barbuy2006}
{Barbuy}, B., {Bica}, E., {Ortolani}, S., \& {Bonatto}, C. 2006, \aap, 449,
  1019

\bibitem[{{Barbuy} {et~al.}(2018){Barbuy}, {Muniz}, {Ortolani}, {Ernandes},
  {Dias}, {Saviane}, {Kerber}, {Bica}, {P{\'e}rez-Villegas}, {Rossi}, \&
  {Held}}]{Barbuy2018}
{Barbuy}, B., {Muniz}, L., {Ortolani}, S., {et~al.} 2018, \aap, 619, A178

\bibitem[{{Barmby} {et~al.}(2000){Barmby}, {Huchra}, {Brodie}, {Forbes},
  {Schroder}, \& {Grillmair}}]{Barmby2000}
{Barmby}, P., {Huchra}, J.~P., {Brodie}, J.~P., {et~al.} 2000, \aj, 119, 727

\bibitem[{{Bassino} \& {Muzzio}(1995)}]{Bassino1995}
{Bassino}, L.~P. \& {Muzzio}, J.~C. 1995, The Observatory, 115, 256

\bibitem[{{Baumgardt} {et~al.}(2020){Baumgardt}, {Sollima}, \&
  {Hilker}}]{Baumgardt2020}
{Baumgardt}, H., {Sollima}, A., \& {Hilker}, M. 2020, \pasa, 37, e046

\bibitem[{{Bellazzini} {et~al.}(2006){Bellazzini}, {Correnti}, {Ferraro},
  {Monaco}, \& {Montegriffo}}]{Bellazzini2006a}
{Bellazzini}, M., {Correnti}, M., {Ferraro}, F.~R., {Monaco}, L., \&
  {Montegriffo}, P. 2006, \aap, 446, L1

\bibitem[{{Bellazzini} {et~al.}(2020){Bellazzini}, {Ibata}, {Malhan}, {Martin},
  {Famaey}, \& {Thomas}}]{Bellazzini2020}
{Bellazzini}, M., {Ibata}, R., {Malhan}, K., {et~al.} 2020, \aap, 636, A107

\bibitem[{{Bellazzini} {et~al.}(2008){Bellazzini}, {Ibata}, {Chapman},
  {Mackey}, {Monaco}, {Irwin}, {Martin}, {Lewis}, \&
  {Dalessandro}}]{Bellazzini2008}
{Bellazzini}, M., {Ibata}, R.~A., {Chapman}, S.~C., {et~al.} 2008, \aj, 136,
  1147

\bibitem[{{Belokurov} {et~al.}(2014){Belokurov}, {Koposov}, {Evans},
  {Pe{\~n}arrubia}, {Irwin}, {Smith}, {Lewis}, {Gieles}, {Wilkinson},
  {Gilmore}, {Olszewski}, \& {Niederste-Ostholt}}]{Belokurov2014}
{Belokurov}, V., {Koposov}, S.~E., {Evans}, N.~W., {et~al.} 2014, \mnras, 437,
  116

\bibitem[{Belokurov {et~al.}(2006)Belokurov, Zucker, Evans, Gilmore, Vidrih,
  Bramich, Newberg, Wyse, Irwin, Fellhauer, Hewett, Walton, Wilkinson, Cole,
  Yanny, Rockosi, Beers, Bell, Brinkmann, Ivezi{\'{c}}, \&
  Lupton}]{Belokurov_2006}
Belokurov, V., Zucker, D.~B., Evans, N.~W., {et~al.} 2006, ApJ, 642, L137

\bibitem[{Besla {et~al.}(2010)Besla, Kallivayalil, Hernquist, van~der Marel,
  Cox, \& Kere{\v{s}}}]{Besla_2010}
Besla, G., Kallivayalil, N., Hernquist, L., {et~al.} 2010, ApJ, 721, L97

\bibitem[{{Bressan} {et~al.}(2012){Bressan}, {Marigo}, {Girardi}, {Salasnich},
  {Dal Cero}, {Rubele}, \& {Nanni}}]{Bressan2012}
{Bressan}, A., {Marigo}, P., {Girardi}, L., {et~al.} 2012, \mnras, 427, 127

\bibitem[{{Bruzual} \& {Charlot}(2003)}]{Bruzual2003}
{Bruzual}, G. \& {Charlot}, S. 2003, \mnras, 344, 1000

\bibitem[{{Chou} {et~al.}(2007){Chou}, {Majewski}, {Cunha}, {Smith},
  {Patterson}, {Mart{\'\i}nez-Delgado}, {Law}, {Crane}, {Mu{\~n}oz}, {Garcia
  L{\'o}pez}, {Geisler}, \& {Skrutskie}}]{Chou2007}
{Chou}, M.-Y., {Majewski}, S.~R., {Cunha}, K., {et~al.} 2007, \apj, 670, 346

\bibitem[{{Christian} \& {Friel}(1992)}]{Christian1992}
{Christian}, C.~A. \& {Friel}, E.~D. 1992, \aj, 103, 142

\bibitem[{{Clementini} {et~al.}(2005){Clementini}, {Gratton}, {Bragaglia},
  {Ripepi}, {Martinez Fiorenzano}, {Held}, \& {Carretta}}]{Clementini2005}
{Clementini}, G., {Gratton}, R.~G., {Bragaglia}, A., {et~al.} 2005, \apjl, 630,
  L145

\bibitem[{{Cohen} {et~al.}(2007){Cohen}, {Hsieh}, {Metchev}, {Djorgovski}, \&
  {Malkan}}]{Cohen2007}
{Cohen}, J.~G., {Hsieh}, S., {Metchev}, S., {Djorgovski}, S.~G., \& {Malkan},
  M. 2007, \aj, 133, 99

\bibitem[{Cohen {et~al.}(2015)Cohen, Hempel, Mauro, Geisler, Alonso-Garcia, \&
  Kinemuchi}]{Cohen_2015}
Cohen, R.~E., Hempel, M., Mauro, F., {et~al.} 2015, AJ, 150, 176

\bibitem[{Cohen {et~al.}(2014)Cohen, Hickox, Wegner, Einasto, \&
  Vennik}]{Cohen_2014}
Cohen, S.~A., Hickox, R.~C., Wegner, G.~A., Einasto, M., \& Vennik, J. 2014,
  ApJ, 783, 136

\bibitem[{{Colucci} {et~al.}(2011){Colucci}, {Bernstein}, {Cameron}, \&
  {McWilliam}}]{Colucci2011}
{Colucci}, J.~E., {Bernstein}, R.~A., {Cameron}, S.~A., \& {McWilliam}, A.
  2011, \apj, 735, 55

\bibitem[{Conroy \& Gunn(2010)}]{Conroy_2010}
Conroy, C. \& Gunn, J.~E. 2010, \aj, 712, 833

\bibitem[{{Cross} {et~al.}(2012){Cross}, {Collins}, {Mann}, {Read}, {Sutorius},
  {Blake}, {Holliman}, {Hambly}, {Emerson}, {Lawrence}, \&
  {Noddle}}]{Cross2012}
{Cross}, N.~J.~G., {Collins}, R.~S., {Mann}, R.~G., {et~al.} 2012, \aap, 548,
  A119

\bibitem[{De~Grijs {et~al.}(2005)De~Grijs, Wilkinson, \&
  Tadhunter}]{DeGrijs2005}
De~Grijs, R., Wilkinson, M.~I., \& Tadhunter, C.~N. 2005, MNRAS, 361, 311

\bibitem[{Di~Benedetto(2008)}]{DiBenedetto2008}
Di~Benedetto, G.~P. 2008, MNRAS, 390, 1762

\bibitem[{Dierickx \& Loeb(2017)}]{Dierickx_2017}
Dierickx, M. I.~P. \& Loeb, A. 2017, ApJ, 847, 42

\bibitem[{{Einasto} {et~al.}(2012){Einasto}, {Vennik}, {Nurmi}, {Tempel},
  {Ahvensalmi}, {Tago}, {Liivam{\"a}gi}, {Saar}, {Hein{\"a}m{\"a}ki},
  {Einasto}, \& {Mart{\'\i}nez}}]{Einasto2012}
{Einasto}, M., {Vennik}, J., {Nurmi}, P., {et~al.} 2012, \aap, 540, A123

\bibitem[{{Emerson} \& {Sutherland}(2010)}]{Emerson2010}
{Emerson}, J. \& {Sutherland}, W. 2010, The Messenger, 139, 2

\bibitem[{Fall \& Zhang(2001)}]{Fall_2001}
Fall, S.~M. \& Zhang, Q. 2001, \aj, 561, 751

\bibitem[{Fan {et~al.}(2008)Fan, Ma, De~Grijs, \& Zhou}]{Fan2008}
Fan, Z., Ma, J., De~Grijs, R., \& Zhou, X. 2008, MNRAS, 385, 1973

\bibitem[{Ferraro {et~al.}(2000)Ferraro, Montegriffo, Origlia, \&
  Pecci}]{Ferraro_2000}
Ferraro, F.~R., Montegriffo, P., Origlia, L., \& Pecci, F.~F. 2000, The
  Astronomical Journal, 119, 1282

\bibitem[{Forbes(2020)}]{Forbes2020}
Forbes, D.~A. 2020, MNRAS, 493, 847

\bibitem[{Forbes \& Bridges(2010)}]{Forbes_Bridges2010}
Forbes, D.~A. \& Bridges, T. 2010, MNRAS, 404, 1203

\bibitem[{{Freedman} {et~al.}(2020){Freedman}, {Madore}, {Hoyt}, {Jang},
  {Beaton}, {Lee}, {Monson}, {Neeley}, \& {Rich}}]{Freedman2020}
{Freedman}, W.~L., {Madore}, B.~F., {Hoyt}, T., {et~al.} 2020, \apj, 891, 57

\bibitem[{{Gaia Collaboration} {et~al.}(2021){Gaia Collaboration}, {Brown},
  {Vallenari}, {Prusti}, {de Bruijne}, {Babusiaux}, {Biermann}, {Creevey},
  {Evans}, {Eyer}, {Hutton}, {Jansen}, {Jordi}, {Klioner}, {Lammers},
  {Lindegren}, {Luri}, {Mignard}, {Panem}, {Pourbaix}, {Randich}, {Sartoretti},
  {Soubiran}, {Walton}, {Arenou}, {Bailer-Jones}, {Bastian}, {Cropper},
  {Drimmel}, {Katz}, {Lattanzi}, {van Leeuwen}, {Bakker}, {Cacciari},
  {Casta{\~n}eda}, {De Angeli}, {Ducourant}, {Fabricius}, {Fouesneau},
  {Fr{\'e}mat}, {Guerra}, {Guerrier}, {Guiraud}, {Jean-Antoine Piccolo},
  {Masana}, {Messineo}, {Mowlavi}, {Nicolas}, {Nienartowicz}, {Pailler},
  {Panuzzo}, {Riclet}, {Roux}, {Seabroke}, {Sordo}, {Tanga}, {Th{\'e}venin},
  {Gracia-Abril}, {Portell}, {Teyssier}, {Altmann}, {Andrae}, {Bellas-Velidis},
  {Benson}, {Berthier}, {Blomme}, {Brugaletta}, {Burgess}, {Busso}, {Carry},
  {Cellino}, {Cheek}, {Clementini}, {Damerdji}, {Davidson}, {Delchambre},
  {Dell'Oro}, {Fern{\'a}ndez-Hern{\'a}ndez}, {Galluccio}, {Garc{\'\i}a-Lario},
  {Garcia-Reinaldos}, {Gonz{\'a}lez-N{\'u}{\~n}ez}, {Gosset}, {Haigron},
  {Halbwachs}, {Hambly}, {Harrison}, {Hatzidimitriou}, {Heiter},
  {Hern{\'a}ndez}, {Hestroffer}, {Hodgkin}, {Holl}, {Jan{\ss}en}, {Jevardat de
  Fombelle}, {Jordan}, {Krone-Martins}, {Lanzafame}, {L{\"o}ffler}, {Lorca},
  {Manteiga}, {Marchal}, {Marrese}, {Moitinho}, {Mora}, {Muinonen}, {Osborne},
  {Pancino}, {Pauwels}, {Petit}, {Recio-Blanco}, {Richards}, {Riello},
  {Rimoldini}, {Robin}, {Roegiers}, {Rybizki}, {Sarro}, {Siopis}, {Smith},
  {Sozzetti}, {Ulla}, {Utrilla}, {van Leeuwen}, {van Reeven}, {Abbas}, {Abreu
  Aramburu}, {Accart}, {Aerts}, {Aguado}, {Ajaj}, {Altavilla}, {{\'A}lvarez},
  {{\'A}lvarez Cid-Fuentes}, {Alves}, {Anderson}, {Anglada Varela}, {Antoja},
  {Audard}, {Baines}, {Baker}, {Balaguer-N{\'u}{\~n}ez}, {Balbinot}, {Balog},
  {Barache}, {Barbato}, {Barros}, {Barstow}, {Bartolom{\'e}}, {Bassilana},
  {Bauchet}, {Baudesson-Stella}, {Becciani}, {Bellazzini}, {Bernet}, {Bertone},
  {Bianchi}, {Blanco-Cuaresma}, {Boch}, {Bombrun}, {Bossini}, {Bouquillon},
  {Bragaglia}, {Bramante}, {Breedt}, {Bressan}, {Brouillet}, {Bucciarelli},
  {Burlacu}, {Busonero}, {Butkevich}, {Buzzi}, {Caffau}, {Cancelliere},
  {C{\'a}novas}, {Cantat-Gaudin}, {Carballo}, {Carlucci}, {Carnerero},
  {Carrasco}, {Casamiquela}, {Castellani}, {Castro-Ginard}, {Castro Sampol},
  {Chaoul}, {Charlot}, {Chemin}, {Chiavassa}, {Cioni}, {Comoretto}, {Cooper},
  {Cornez}, {Cowell}, {Crifo}, {Crosta}, {Crowley}, {Dafonte}, {Dapergolas},
  {David}, {David}, {de Laverny}, {De Luise}, {De March}, {De Ridder}, {de
  Souza}, {de Teodoro}, {de Torres}, {del Peloso}, {del Pozo}, {Delbo},
  {Delgado}, {Delgado}, {Delisle}, {Di Matteo}, {Diakite}, {Diener},
  {Distefano}, {Dolding}, {Eappachen}, {Edvardsson}, {Enke}, {Esquej}, {Fabre},
  {Fabrizio}, {Faigler}, {Fedorets}, {Fernique}, {Fienga}, {Figueras},
  {Fouron}, {Fragkoudi}, {Fraile}, {Franke}, {Gai}, {Garabato},
  {Garcia-Gutierrez}, {Garc{\'\i}a-Torres}, {Garofalo}, {Gavras}, {Gerlach},
  {Geyer}, {Giacobbe}, {Gilmore}, {Girona}, {Giuffrida}, {Gomel}, {Gomez},
  {Gonzalez-Santamaria}, {Gonz{\'a}lez-Vidal}, {Granvik},
  {Guti{\'e}rrez-S{\'a}nchez}, {Guy}, {Hauser}, {Haywood}, {Helmi}, {Hidalgo},
  {Hilger}, {H{\l}adczuk}, {Hobbs}, {Holland}, {Huckle}, {Jasniewicz},
  {Jonker}, {Juaristi Campillo}, {Julbe}, {Karbevska}, {Kervella}, {Khanna},
  {Kochoska}, {Kontizas}, {Kordopatis}, {Korn}, {Kostrzewa-Rutkowska},
  {Kruszy{\'n}ska}, {Lambert}, {Lanza}, {Lasne}, {Le Campion}, {Le Fustec},
  {Lebreton}, {Lebzelter}, {Leccia}, {Leclerc}, {Lecoeur-Taibi}, {Liao},
  {Licata}, {Lindstr{\o}m}, {Lister}, {Livanou}, {Lobel}, {Madrero Pardo},
  {Managau}, {Mann}, {Marchant}, {Marconi}, {Marcos Santos}, {Marinoni},
  {Marocco}, {Marshall}, {Martin Polo}, {Mart{\'\i}n-Fleitas}, {Masip},
  {Massari}, {Mastrobuono-Battisti}, {Mazeh}, {McMillan}, {Messina},
  {Michalik}, {Millar}, {Mints}, {Molina}, {Molinaro}, {Moln{\'a}r},
  {Montegriffo}, {Mor}, {Morbidelli}, {Morel}, {Morris}, {Mulone}, {Munoz},
  {Muraveva}, {Murphy}, {Musella}, {Noval}, {Ord{\'e}novic}, {Orr{\`u}},
  {Osinde}, {Pagani}, {Pagano}, {Palaversa}, {Palicio}, {Panahi}, {Pawlak},
  {Pe{\~n}alosa Esteller}, {Penttil{\"a}}, {Piersimoni}, {Pineau}, {Plachy},
  {Plum}, {Poggio}, {Poretti}, {Poujoulet}, {Pr{\v{s}}a}, {Pulone}, {Racero},
  {Ragaini}, {Rainer}, {Raiteri}, {Rambaux}, {Ramos}, {Ramos-Lerate}, {Re
  Fiorentin}, {Regibo}, {Reyl{\'e}}, {Ripepi}, {Riva}, {Rixon}, {Robichon},
  {Robin}, {Roelens}, {Rohrbasser}, {Romero-G{\'o}mez}, {Rowell}, {Royer},
  {Rybicki}, {Sadowski}, {Sagrist{\`a} Sell{\'e}s}, {Sahlmann}, {Salgado},
  {Salguero}, {Samaras}, {Sanchez Gimenez}, {Sanna}, {Santove{\~n}a},
  {Sarasso}, {Schultheis}, {Sciacca}, {Segol}, {Segovia}, {S{\'e}gransan},
  {Semeux}, {Shahaf}, {Siddiqui}, {Siebert}, {Siltala}, {Slezak}, {Smart},
  {Solano}, {Solitro}, {Souami}, {Souchay}, {Spagna}, {Spoto}, {Steele},
  {Steidelm{\"u}ller}, {Stephenson}, {S{\"u}veges}, {Szabados}, {Szegedi-Elek},
  {Taris}, {Tauran}, {Taylor}, {Teixeira}, {Thuillot}, {Tonello}, {Torra},
  {Torra}, {Turon}, {Unger}, {Vaillant}, {van Dillen}, {Vanel}, {Vecchiato},
  {Viala}, {Vicente}, {Voutsinas}, {Weiler}, {Wevers}, {Wyrzykowski}, {Yoldas},
  {Yvard}, {Zhao}, {Zorec}, {Zucker}, {Zurbach}, \&
  {Zwitter}}]{GaiaCollaboration2020}
{Gaia Collaboration}, {Brown}, A.~G.~A., {Vallenari}, A., {et~al.} 2021, \aap,
  649, A1

\bibitem[{{Galleti} {et~al.}(2006){Galleti}, {Federici}, {Bellazzini},
  {Buzzoni}, \& {Fusi Pecci}}]{Galleti2006}
{Galleti}, S., {Federici}, L., {Bellazzini}, M., {Buzzoni}, A., \& {Fusi
  Pecci}, F. 2006, \aap, 456, 985

\bibitem[{{Galleti} {et~al.}(2014){Galleti}, {Federici}, {Bellazzini}, {Fusi
  Pecci}, {Macrina}, \& {Buzzoni}}]{Galleti2014}
{Galleti}, S., {Federici}, L., {Bellazzini}, M., {et~al.} 2014, VizieR Online
  Data Catalog, V/143

\bibitem[{{Garro} {et~al.}(2020){Garro}, {Minniti}, {G{\'o}mez},
  {Alonso-Garc{\'\i}a}, {Barb{\'a}}, {Barbuy}, {Clari{\'a}}, {Chen{\'e}},
  {Dias}, {Hempel}, {Ivanov}, {Lucas}, {Majaess}, {Mauro}, {Moni Bidin},
  {Palma}, {Pullen}, {Saito}, {Smith}, {Surot}, {Ram{\'\i}rez Alegr{\'\i}a},
  {Rejkuba}, {Ripepi}, \& {Fern{\'a}ndez Trincado}}]{Garro2020}
{Garro}, E.~R., {Minniti}, D., {G{\'o}mez}, M., {et~al.} 2020, \aap, 642, L19

\bibitem[{{Garro} {et~al.}(2021){Garro}, {Minniti}, {G{\'o}mez},
  {Alonso-Garc{\'\i}a}, {Palma}, {Smith}, \& {Ripepi}}]{Garro2021a}
{Garro}, E.~R., {Minniti}, D., {G{\'o}mez}, M., {et~al.} 2021, \aap, 649, A86

\bibitem[{{Gnedin} \& {Ostriker}(1997)}]{Gnedin1997}
{Gnedin}, O.~Y. \& {Ostriker}, J.~P. 1997, \apj, 474, 223

\bibitem[{{Gratton} {et~al.}(2003){Gratton}, {Bragaglia}, {Carretta},
  {Clementini}, {Desidera}, {Grundahl}, \& {Lucatello}}]{Gratton2003}
{Gratton}, R.~G., {Bragaglia}, A., {Carretta}, E., {et~al.} 2003, \aap, 408,
  529

\bibitem[{{Grcevich} \& {Putman}(2009)}]{Grcevich2009}
{Grcevich}, J. \& {Putman}, M.~E. 2009, \apj, 696, 385

\bibitem[{{Haghi} {et~al.}(2017){Haghi}, {Khalaj}, {Hasani Zonoozi}, \&
  {Kroupa}}]{Haghi2017}
{Haghi}, H., {Khalaj}, P., {Hasani Zonoozi}, A., \& {Kroupa}, P. 2017, \apj,
  839, 60

\bibitem[{{Hamanowicz} {et~al.}(2016){Hamanowicz}, {Pietrukowicz}, {Udalski},
  {Mr{\'o}z}, {Soszy{\'n}ski}, {Szyma{\'n}ski}, {Skowron}, {Poleski},
  {Wyrzykowski}, {Koz{\l}owski}, {Pawlak}, \& {Ulaczyk}}]{Hamanowicz2016}
{Hamanowicz}, A., {Pietrukowicz}, P., {Udalski}, A., {et~al.} 2016, \actaa, 66,
  197

\bibitem[{{Harris}(1991)}]{Harris_1991}
{Harris}, W.~E. 1991, \araa, 29, 543

\bibitem[{{Harris}(1996)}]{Harris1996}
{Harris}, W.~E. 1996, \aj, 112, 1487

\bibitem[{{Helmi} {et~al.}(2018){Helmi}, {Babusiaux}, {Koppelman}, {Massari},
  {Veljanoski}, \& {Brown}}]{Helmi2018}
{Helmi}, A., {Babusiaux}, C., {Koppelman}, H.~H., {et~al.} 2018, \nat, 563, 85

\bibitem[{{Helmi} {et~al.}(1999){Helmi}, {White}, {de Zeeuw}, \&
  {Zhao}}]{Helmi1999}
{Helmi}, A., {White}, S. D.~M., {de Zeeuw}, P.~T., \& {Zhao}, H. 1999, \nat,
  402, 53

\bibitem[{{Hilker} {et~al.}(2020){Hilker}, {Baumgardt}, {Sollima}, \&
  {Bellini}}]{Hilker2020}
{Hilker}, M., {Baumgardt}, H., {Sollima}, A., \& {Bellini}, A. 2020, in Star
  Clusters: From the Milky Way to the Early Universe, ed. A.~{Bragaglia},
  M.~{Davies}, A.~{Sills}, \& E.~{Vesperini}, Vol. 351, 451--454

\bibitem[{{Hollyhead} {et~al.}(2019){Hollyhead}, {Martocchia}, {Lardo},
  {Bastian}, {Kacharov}, {Niederhofer}, {Cabrera-Ziri}, {Dalessandro},
  {Mucciarelli}, {Salaris}, \& {Usher}}]{Hollyhead2019}
{Hollyhead}, K., {Martocchia}, S., {Lardo}, C., {et~al.} 2019, \mnras, 484,
  4718

\bibitem[{{Horta} {et~al.}(2021){Horta}, {Hughes}, {Pfeffer}, {Bastian},
  {Kruijssen}, {Reina-Campos}, \& {Crain}}]{Horta2021}
{Horta}, D., {Hughes}, M.~E., {Pfeffer}, J.~L., {et~al.} 2021, \mnras, 500,
  4768

\bibitem[{Hughes {et~al.}(2019)Hughes, Pfeffer, Martig, Bastian, Crain,
  Kruijssen, \& Reina-Campos}]{Hughes2019}
Hughes, M.~E., Pfeffer, J., Martig, M., {et~al.} 2019, MNRAS, 482, 2795

\bibitem[{Huxor {et~al.}(2014)Huxor, Mackey, Ferguson, Irwin, Martin, Tanvir,
  Veljanoski, McConnachie, Fishlock, Ibata, \& Lewis}]{Huxor2014}
Huxor, A.~P., Mackey, A.~D., Ferguson, A. M.~N., {et~al.} 2014, MNRAS, 442,
  2165

\bibitem[{{Ibata} {et~al.}(2004){Ibata}, {Chapman}, {Ferguson}, {Irwin},
  {Lewis}, \& {McConnachie}}]{Ibata2004}
{Ibata}, R., {Chapman}, S., {Ferguson}, A.~M.~N., {et~al.} 2004, \mnras, 351,
  117

\bibitem[{{Ibata} {et~al.}(2001){Ibata}, {Irwin}, {Lewis}, {Ferguson}, \&
  {Tanvir}}]{Ibata2001}
{Ibata}, R., {Irwin}, M., {Lewis}, G., {Ferguson}, A. M.~N., \& {Tanvir}, N.
  2001, \nat, 412, 49

\bibitem[{{Ibata} {et~al.}(1994){Ibata}, {Gilmore}, \& {Irwin}}]{Ibata1994}
{Ibata}, R.~A., {Gilmore}, G., \& {Irwin}, M.~J. 1994, \nat, 370, 194

\bibitem[{{Irwin} {et~al.}(2004){Irwin}, {Lewis}, {Hodgkin}, {Bunclark},
  {Evans}, {McMahon}, {Emerson}, {Stewart}, \& {Beard}}]{Irwin2004}
{Irwin}, M.~J., {Lewis}, J., {Hodgkin}, S., {et~al.} 2004, in Society of
  Photo-Optical Instrumentation Engineers (SPIE) Conference Series, Vol. 5493,
  Optimizing Scientific Return for Astronomy through Information Technologies,
  ed. P.~J. {Quinn} \& A.~{Bridger}, 411--422

\bibitem[{Jeon {et~al.}(2014)Jeon, Nemec, Walker, \& Kunder}]{Jeon_2014}
Jeon, Y.-B., Nemec, J.~M., Walker, A.~R., \& Kunder, A.~M. 2014, AJ, 147, 155

\bibitem[{Kallivayalil {et~al.}(2006)Kallivayalil, van~der Marel, \&
  Alcock}]{Kallivayalil_2006}
Kallivayalil, N., van~der Marel, R.~P., \& Alcock, C. 2006, ApJ, 652, 1213

\bibitem[{Kallivayalil {et~al.}(2013)Kallivayalil, van~der Marel, Besla,
  Anderson, \& Alcock}]{Kallivayalil_2013}
Kallivayalil, N., van~der Marel, R.~P., Besla, G., Anderson, J., \& Alcock, C.
  2013, ApJ, 764, 161

\bibitem[{{Kerber} {et~al.}(2018){Kerber}, {Nardiello}, {Ortolani}, {Barbuy},
  {Bica}, {Cassisi}, {Libralato}, \& {Vieira}}]{Kerber2018}
{Kerber}, L.~O., {Nardiello}, D., {Ortolani}, S., {et~al.} 2018, \apj, 853, 15

\bibitem[{{Kerber} {et~al.}(2007){Kerber}, {Santiago}, \&
  {Brocato}}]{Kerber2007}
{Kerber}, L.~O., {Santiago}, B.~X., \& {Brocato}, E. 2007, \aap, 462, 139

\bibitem[{{Kharchenko} {et~al.}(2016){Kharchenko}, {Piskunov}, {Schilbach},
  {R{\"o}ser}, \& {Scholz}}]{Kharchenko2016}
{Kharchenko}, N.~V., {Piskunov}, A.~E., {Schilbach}, E., {R{\"o}ser}, S., \&
  {Scholz}, R.~D. 2016, \aap, 585, A101

\bibitem[{{Koposov} {et~al.}(2007){Koposov}, {de Jong}, {Belokurov}, {Rix},
  {Zucker}, {Evans}, {Gilmore}, {Irwin}, \& {Bell}}]{Koposov2007}
{Koposov}, S., {de Jong}, J.~T.~A., {Belokurov}, V., {et~al.} 2007, \apj, 669,
  337

\bibitem[{Kruijssen {et~al.}(2011)Kruijssen, Pelupessy, Lamers,
  Portegies~Zwart, \& Icke}]{Kruijssen2011}
Kruijssen, J. M.~D., Pelupessy, F.~I., Lamers, H. J. G. L.~M., Portegies~Zwart,
  S.~F., \& Icke, V. 2011, MNRAS, 414, 1339

\bibitem[{{Kruijssen} {et~al.}(2020){Kruijssen}, {Pfeffer}, {Chevance},
  {Bonaca}, {Trujillo-Gomez}, {Bastian}, {Reina-Campos}, {Crain}, \&
  {Hughes}}]{Kruijssen2020}
{Kruijssen}, J.~M.~D., {Pfeffer}, J.~L., {Chevance}, M., {et~al.} 2020, \mnras,
  498, 2472

\bibitem[{{Kruijssen} {et~al.}(2019){Kruijssen}, {Pfeffer}, {Reina-Campos},
  {Crain}, \& {Bastian}}]{Kruijssen2019}
{Kruijssen}, J.~M.~D., {Pfeffer}, J.~L., {Reina-Campos}, M., {Crain}, R.~A., \&
  {Bastian}, N. 2019, \mnras, 486, 3180

\bibitem[{Lamers {et~al.}(2010)Lamers, Baumgardt, \& Gieles}]{Lamers2010}
Lamers, H. J. G. L.~M., Baumgardt, H., \& Gieles, M. 2010, Monthly Notices of
  the Royal Astronomical Society, 409, 305

\bibitem[{Laporte {et~al.}(2018)Laporte, Johnston, Gómez, Garavito-Camargo, \&
  Besla}]{Laporte2018}
Laporte, C. F.~P., Johnston, K.~V., Gómez, F.~A., Garavito-Camargo, N., \&
  Besla, G. 2018, MNRAS, 481, 286

\bibitem[{Larsen(2002)}]{Larsen_2002}
Larsen, S.~S. 2002, AJ, 124, 1393

\bibitem[{Law \& Majewski(2010)}]{Law_2010}
Law, D.~R. \& Majewski, S.~R. 2010, ApJ, 714, 229

\bibitem[{{Layden} \& {Sarajedini}(2000)}]{Layden2000}
{Layden}, A.~C. \& {Sarajedini}, A. 2000, \aj, 119, 1760

\bibitem[{Leaman {et~al.}(2013)Leaman, VandenBerg, \& Mendel}]{Leaman2013}
Leaman, R., VandenBerg, D.~A., \& Mendel, J.~T. 2013, MNRAS, 436, 122

\bibitem[{{Lyubenova} {et~al.}(2010){Lyubenova}, {Kuntschner}, {Rejkuba},
  {Silva}, {Kissler-Patig}, {Tacconi-Garman}, \& {Larsen}}]{Lyubenova2010}
{Lyubenova}, M., {Kuntschner}, H., {Rejkuba}, M., {et~al.} 2010, \aap, 510, A19

\bibitem[{Mackey \& Gilmore(2004)}]{Mackey2004}
Mackey, A.~D. \& Gilmore, G.~F. 2004, MNRAS, 352, 153

\bibitem[{Mackey \& Van Den~Bergh(2005)}]{Mackey2005}
Mackey, A.~D. \& Van Den~Bergh, S. 2005, MNRAS, 360, 631

\bibitem[{Madrid {et~al.}(2017)Madrid, Leigh, Hurley, \& Giersz}]{Madrid2017}
Madrid, J.~P., Leigh, N. W.~C., Hurley, J.~R., \& Giersz, M. 2017, Monthly
  Notices of the Royal Astronomical Society, 470, 1729

\bibitem[{{Majewski} {et~al.}(2003){Majewski}, {Skrutskie}, {Weinberg}, \&
  {Ostheimer}}]{Majewski2003}
{Majewski}, S.~R., {Skrutskie}, M.~F., {Weinberg}, M.~D., \& {Ostheimer}, J.~C.
  2003, \apj, 599, 1082

\bibitem[{{Marigo} {et~al.}(2017){Marigo}, {Girardi}, {Bressan}, {Rosenfield},
  {Aringer}, {Chen}, {Dussin}, {Nanni}, {Pastorelli}, {Rodrigues}, {Trabucchi},
  {Bladh}, {Dalcanton}, {Groenewegen}, {Montalb{\'a}n}, \& {Wood}}]{Marigo2017}
{Marigo}, P., {Girardi}, L., {Bressan}, A., {et~al.} 2017, \apj, 835, 77

\bibitem[{Martin {et~al.}(2014)Martin, Ibata, Rich, Collins, Fardal, Irwin,
  Lewis, McConnachie, Babul, Bate, Chapman, Conn, Crnojevi{\'{c}}, Ferguson,
  Mackey, Navarro, Pe{\~{n}}arrubia, Tanvir, \& Valls-Gabaud}]{Martin_2014}
Martin, N.~F., Ibata, R.~A., Rich, R.~M., {et~al.} 2014, ApJ, 787, 19

\bibitem[{{Martocchia} {et~al.}(2019){Martocchia}, {Dalessandro}, {Lardo},
  {Cabrera-Ziri}, {Bastian}, {Kozhurina-Platais}, {Salaris}, {Chantereau},
  {Geisler}, {Hilker}, {Kacharov}, {Larsen}, {Mucciarelli}, {Niederhofer},
  {Platais}, \& {Usher}}]{Martocchia2019}
{Martocchia}, S., {Dalessandro}, E., {Lardo}, C., {et~al.} 2019, \mnras, 487,
  5324

\bibitem[{{Massari} {et~al.}(2019){Massari}, {Koppelman}, \&
  {Helmi}}]{Massari2019}
{Massari}, D., {Koppelman}, H.~H., \& {Helmi}, A. 2019, \aap, 630, L4

\bibitem[{{McConnachie} {et~al.}(2005){McConnachie}, {Irwin}, {Ferguson},
  {Ibata}, {Lewis}, \& {Tanvir}}]{McConnachie2005}
{McConnachie}, A.~W., {Irwin}, M.~J., {Ferguson}, A.~M.~N., {et~al.} 2005,
  \mnras, 356, 979

\bibitem[{{McDonald} {et~al.}(2014){McDonald}, {Zijlstra}, {Sloan}, {Kerins},
  {Lagadec}, \& {Minniti}}]{McDonald2014}
{McDonald}, I., {Zijlstra}, A.~A., {Sloan}, G.~C., {et~al.} 2014, \mnras, 439,
  2618

\bibitem[{McDonald {et~al.}(2013)McDonald, Zijlstra, Sloan, Kerins, Lagadec,
  Minniti, Santucho, Gurovich, \& Domínguez~Romero}]{McDonald2013}
McDonald, I., Zijlstra, A.~A., Sloan, G.~C., {et~al.} 2013, MNRAS, 436, 413

\bibitem[{{McLaughlin} \& {van der Marel}(2005)}]{McLaughlin2005}
{McLaughlin}, D.~E. \& {van der Marel}, R.~P. 2005, \apjs, 161, 304

\bibitem[{{Minniti}(2018)}]{Minniti2018}
{Minniti}, D. 2018, in The Vatican Observatory, Castel Gandolfo: 80th
  Anniversary Celebration, ed. G.~{Gionti} \& J.-B. {Kikwaya Eluo}, Vol.~51, 63

\bibitem[{Minniti {et~al.}(2017{\natexlab{a}})Minniti, Geisler,
  Alonso-Garc{\'{\i}}a, Palma, Beam{\'{\i}}n, Borissova, Catelan, Clari{\'{a}},
  Cohen, Ramos, Dias, Fern{\'{a}}ndez-Trincado, G{\'{o}}mez, Hempel, Ivanov,
  Kurtev, Lucas, Moni-Bidin, Pullen, Alegr{\'{\i}}a, Saito, \&
  Valenti}]{Minniti_2017a}
Minniti, D., Geisler, D., Alonso-Garc{\'{\i}}a, J., {et~al.}
  2017{\natexlab{a}}, \aj, 849, L24

\bibitem[{{Minniti} {et~al.}(2021{\natexlab{a}}){Minniti}, {G{\'o}mez},
  {Alonso-Garc{\'\i}a}, {Saito}, \& {Garro}}]{Minniti2021_second}
{Minniti}, D., {G{\'o}mez}, M., {Alonso-Garc{\'\i}a}, J., {Saito}, R.~K., \&
  {Garro}, E.~R. 2021{\natexlab{a}}, \aap, 650, L21

\bibitem[{{Minniti} {et~al.}(2011){Minniti}, {Hempel}, {Toledo}, {Ivanov},
  {Alonso-Garc{\'\i}a}, {Saito}, {Catelan}, {Geisler}, {Jord{\'a}n},
  {Borissova}, {Zoccali}, {Kurtev}, {Carraro}, {Barbuy}, {Clari{\'a}},
  {Rejkuba}, {Emerson}, \& {Moni Bidin}}]{Minniti2011}
{Minniti}, D., {Hempel}, M., {Toledo}, I., {et~al.} 2011, \aap, 527, A81

\bibitem[{{Minniti} {et~al.}(2010){Minniti}, {Lucas}, {Emerson}, {Saito},
  {Hempel}, {Pietrukowicz}, {Ahumada}, {Alonso}, {Alonso-Garcia}, {Arias},
  {Bandyopadhyay}, {Barb{\'a}}, {Barbuy}, {Bedin}, {Bica}, {Borissova},
  {Bronfman}, {Carraro}, {Catelan}, {Clari{\'a}}, {Cross}, {de Grijs},
  {D{\'e}k{\'a}ny}, {Drew}, {Fari{\~n}a}, {Feinstein}, {Fern{\'a}ndez
  Laj{\'u}s}, {Gamen}, {Geisler}, {Gieren}, {Goldman}, {Gonzalez}, {Gunthardt},
  {Gurovich}, {Hambly}, {Irwin}, {Ivanov}, {Jord{\'a}n}, {Kerins}, {Kinemuchi},
  {Kurtev}, {L{\'o}pez-Corredoira}, {Maccarone}, {Masetti}, {Merlo},
  {Messineo}, {Mirabel}, {Monaco}, {Morelli}, {Padilla}, {Palma}, {Parisi},
  {Pignata}, {Rejkuba}, {Roman-Lopes}, {Sale}, {Schreiber}, {Schr{\"o}der},
  {Smith}, {}, {Soto}, {Tamura}, {Tappert}, {Thompson}, {Toledo}, {Zoccali}, \&
  {Pietrzynski}}]{Minniti2010}
{Minniti}, D., {Lucas}, P.~W., {Emerson}, J.~P., {et~al.} 2010, \na, 15, 433

\bibitem[{Minniti {et~al.}(2017{\natexlab{b}})Minniti, Palma,
  D{\'{e}}k{\'{a}}ny, Hempel, Rejkuba, Pullen, Alonso-Garc{\'{\i}}a,
  Barb{\'{a}}, Barbuy, Bica, Bonatto, Borissova, Catelan, Carballo-Bello,
  Chene, Clari{\'{a}}, Cohen, Ramos, Dias, Emerson, Froebrich, Buckner,
  Geisler, Gonzalez, Gran, Hagdu, Irwin, Ivanov, Kurtev, Lucas, Majaess, Mauro,
  Moni-Bidin, Navarrete, Alegr{\'{\i}}a, Saito, Valenti, \&
  Zoccali}]{Minniti_2017b}
Minniti, D., Palma, T., D{\'{e}}k{\'{a}}ny, I., {et~al.} 2017{\natexlab{b}},
  \aj, 838, L14

\bibitem[{{Minniti} {et~al.}(2021{\natexlab{b}}){Minniti}, {Ripepi},
  {Fern{\'a}ndez-Trincado}, {Alonso-Garc{\'\i}a}, {Smith}, {Lucas},
  {G{\'o}mez}, {Pullen}, {Garro}, {Vivanco C{\'a}diz}, {Hempel}, {Rejkuba},
  {Saito}, {Palma}, {Clari{\'a}}, {Gregg}, \& {Majaess}}]{Minniti2021_first}
{Minniti}, D., {Ripepi}, V., {Fern{\'a}ndez-Trincado}, J.~G., {et~al.}
  2021{\natexlab{b}}, \aap, 647, L4

\bibitem[{Minniti {et~al.}(2018)Minniti, Schlafly, Palma, Clari{\'{a}}, Hempel,
  Alonso-Garc{\'{\i}}a, Bica, Bonatto, Braga, Clementini, Garofalo,
  G{\'{o}}mez, Ivanov, Lucas, Pullen, Saito, \& Smith}]{Minniti_2018}
Minniti, D., Schlafly, E.~F., Palma, T., {et~al.} 2018, \aj, 866, 12

\bibitem[{{Momany} {et~al.}(2005){Momany}, {Held}, {Saviane}, {Bedin},
  {Gullieuszik}, {Clemens}, {Rizzi}, {Rich}, \& {Kuijken}}]{Momany2005}
{Momany}, Y., {Held}, E.~V., {Saviane}, I., {et~al.} 2005, \aap, 439, 111

\bibitem[{{Monaco} {et~al.}(2005){Monaco}, {Bellazzini}, {Bonifacio},
  {Ferraro}, {Marconi}, {Pancino}, {Sbordone}, \& {Zaggia}}]{Monaco2005b}
{Monaco}, L., {Bellazzini}, M., {Bonifacio}, P., {et~al.} 2005, \aap, 441, 141

\bibitem[{{Monaco} {et~al.}(2004){Monaco}, {Pancino}, {Ferraro}, \&
  {Bellazzini}}]{Monaco2004}
{Monaco}, L., {Pancino}, E., {Ferraro}, F.~R., \& {Bellazzini}, M. 2004,
  \mnras, 349, 1278

\bibitem[{{Mucciarelli} {et~al.}(2017){Mucciarelli}, {Bellazzini}, {Ibata},
  {Romano}, {Chapman}, \& {Monaco}}]{Mucciarelli2017}
{Mucciarelli}, A., {Bellazzini}, M., {Ibata}, R., {et~al.} 2017, \aap, 605, A46

\bibitem[{{Mucciarelli} {et~al.}(2008){Mucciarelli}, {Carretta}, {Origlia}, \&
  {Ferraro}}]{Mucciarelli2008}
{Mucciarelli}, A., {Carretta}, E., {Origlia}, L., \& {Ferraro}, F.~R. 2008,
  \aj, 136, 375

\bibitem[{{Mucciarelli} {et~al.}(2014){Mucciarelli}, {Dalessandro}, {Ferraro},
  {Origlia}, \& {Lanzoni}}]{Mucciarelli2014}
{Mucciarelli}, A., {Dalessandro}, E., {Ferraro}, F.~R., {Origlia}, L., \&
  {Lanzoni}, B. 2014, \apjl, 793, L6

\bibitem[{{Mucciarelli} {et~al.}(2012){Mucciarelli}, {Origlia}, {Ferraro},
  {Bellazzini}, \& {Lanzoni}}]{Mucciarelli2012}
{Mucciarelli}, A., {Origlia}, L., {Ferraro}, F.~R., {Bellazzini}, M., \&
  {Lanzoni}, B. 2012, \apjl, 746, L19

\bibitem[{Myeong {et~al.}(2019)Myeong, Vasiliev, Iorio, Evans, \&
  Belokurov}]{Myeong2019}
Myeong, G.~C., Vasiliev, E., Iorio, G., Evans, N.~W., \& Belokurov, V. 2019,
  MNRAS, 488, 1235

\bibitem[{Nantais {et~al.}(2006)Nantais, Huchra, Barmby, Olsen, \&
  Jarrett}]{Nantais_2006}
Nantais, J.~B., Huchra, J.~P., Barmby, P., Olsen, K. A.~G., \& Jarrett, T.~H.
  2006, AJ, 131, 1416

\bibitem[{Newberg {et~al.}(2002)Newberg, Yanny, Rockosi, Grebel, Rix,
  Brinkmann, Csabai, Hennessy, Hindsley, Ibata, Ivezi{\'{c}}, Lamb, Nash,
  Odenkirchen, Rave, Schneider, Smith, Stolte, \& York}]{Newberg_2002}
Newberg, H.~J., Yanny, B., Rockosi, C., {et~al.} 2002, ApJ, 569, 245

\bibitem[{{Niederste-Ostholt} {et~al.}(2012){Niederste-Ostholt}, {Belokurov},
  \& {Evans}}]{Niederste-Ostholt2012}
{Niederste-Ostholt}, M., {Belokurov}, V., \& {Evans}, N.~W. 2012, \mnras, 422,
  207

\bibitem[{Noël {et~al.}(2013)Noël, Greggio, Renzini, Carollo, \&
  Maraston}]{Noel_2013}
Noël, N. E.~D., Greggio, L., Renzini, A., Carollo, C.~M., \& Maraston, C.
  2013, \aj, 772, 58

\bibitem[{{Ortolani} {et~al.}(1998){Ortolani}, {Bica}, \&
  {Barbuy}}]{Ortolani1998}
{Ortolani}, S., {Bica}, E., \& {Barbuy}, B. 1998, \aaps, 127, 471

\bibitem[{{Ortolani} {et~al.}(2006){Ortolani}, {Bica}, \&
  {Barbuy}}]{Ortolani2006}
{Ortolani}, S., {Bica}, E., \& {Barbuy}, B. 2006, \apjl, 646, L115

\bibitem[{{Palma} {et~al.}(2016){Palma}, {Gramajo}, {Clari{\'a}}, {Lares},
  {Geisler}, \& {Ahumada}}]{Palma2016}
{Palma}, T., {Gramajo}, L.~V., {Clari{\'a}}, J.~J., {et~al.} 2016, \aap, 586,
  A41

\bibitem[{Palma {et~al.}(2019)Palma, Minniti, Alonso-García, Crestani, Netzel,
  Clariá, Saito, Dias, Fernández-Trincado, Kammers, Geisler, Gómez, Hempel,
  \& Pullen}]{Palma2019}
Palma, T., Minniti, D., Alonso-García, J., {et~al.} 2019, Monthly Notices of
  the Royal Astronomical Society, 487, 3140

\bibitem[{Peacock {et~al.}(2010)Peacock, Maccarone, Knigge, Kundu, Waters,
  Zepf, \& Zurek}]{Peacock2010}
Peacock, M.~B., Maccarone, T.~J., Knigge, C., {et~al.} 2010, MNRAS, 402, 803

\bibitem[{{Pessev} {et~al.}(2008){Pessev}, {Goudfrooij}, {Puzia}, \&
  {Chandar}}]{Pessev2008}
{Pessev}, P.~M., {Goudfrooij}, P., {Puzia}, T.~H., \& {Chandar}, R. 2008,
  \mnras, 385, 1535

\bibitem[{{Piatti} {et~al.}(2019){Piatti}, {Alfaro}, \&
  {Cantat-Gaudin}}]{Piatti2019}
{Piatti}, A.~E., {Alfaro}, E.~J., \& {Cantat-Gaudin}, T. 2019, \mnras, 484, L19

\bibitem[{{Piatti} \& {Carballo-Bello}(2020)}]{Piatti_Carballo2020}
{Piatti}, A.~E. \& {Carballo-Bello}, J.~A. 2020, \aap, 637, L2

\bibitem[{{Piatti} \& {Koch}(2018)}]{Piatti2018}
{Piatti}, A.~E. \& {Koch}, A. 2018, \apj, 867, 8

\bibitem[{{Pritzl} {et~al.}(2002){Pritzl}, {Smith}, {Catelan}, \&
  {Sweigart}}]{Pritzl2002}
{Pritzl}, B.~J., {Smith}, H.~A., {Catelan}, M., \& {Sweigart}, A.~V. 2002, \aj,
  124, 949

\bibitem[{{Rejkuba}(2012)}]{Rejkuba2012}
{Rejkuba}, M. 2012, \apss, 341, 195

\bibitem[{Rosenberg {et~al.}(1999)Rosenberg, Saviane, Piotto, \&
  Aparicio}]{Rosenberg_1999}
Rosenberg, A., Saviane, I., Piotto, G., \& Aparicio, A. 1999, The Astronomical
  Journal, 118, 2306

\bibitem[{{Ruiz-Dern} {et~al.}(2018){Ruiz-Dern}, {Babusiaux}, {Arenou},
  {Turon}, \& {Lallement}}]{RuizDern2018}
{Ruiz-Dern}, L., {Babusiaux}, C., {Arenou}, F., {Turon}, C., \& {Lallement}, R.
  2018, \aap, 609, A116

\bibitem[{{Saito} {et~al.}(2012){Saito}, {Hempel}, {Minniti}, {Lucas},
  {Rejkuba}, {Toledo}, {Gonzalez}, {Alonso-Garc{\'\i}a}, {Irwin},
  {Gonzalez-Solares}, {Hodgkin}, {Lewis}, {Cross}, {Ivanov}, {Kerins},
  {Emerson}, {Soto}, {Am{\^o}res}, {Gurovich}, {D{\'e}k{\'a}ny}, {Angeloni},
  {Beamin}, {Catelan}, {Padilla}, {Zoccali}, {Pietrukowicz}, {Moni Bidin},
  {Mauro}, {Geisler}, {Folkes}, {Sale}, {Borissova}, {Kurtev}, {Ahumada},
  {Alonso}, {Adamson}, {Arias}, {Bandyopadhyay}, {Barb{\'a}}, {Barbuy},
  {Baume}, {Bedin}, {Bellini}, {Benjamin}, {Bica}, {Bonatto}, {Bronfman},
  {Carraro}, {Chen{\`e}}, {Clari{\'a}}, {Clarke}, {Contreras}, {Corvill{\'o}n},
  {de Grijs}, {Dias}, {Drew}, {Fari{\~n}a}, {Feinstein},
  {Fern{\'a}ndez-Laj{\'u}s}, {Gamen}, {Gieren}, {Goldman},
  {Gonz{\'a}lez-Fern{\'a}ndez}, {Grand}, {Gunthardt}, {Hambly}, {Hanson},
  {He{\l}miniak}, {Hoare}, {Huckvale}, {Jord{\'a}n}, {Kinemuchi}, {Longmore},
  {L{\'o}pez-Corredoira}, {Maccarone}, {Majaess}, {Mart{\'\i}n}, {Masetti},
  {Mennickent}, {Mirabel}, {Monaco}, {Morelli}, {Motta}, {Palma}, {Parisi},
  {Parker}, {Pe{\~n}aloza}, {Pietrzy{\'n}ski}, {Pignata}, {Popescu}, {Read},
  {Rojas}, {Roman-Lopes}, {Ruiz}, {Saviane}, {Schreiber}, {Schr{\"o}der},
  {Sharma}, {Smith}, {Sodr{\'e}}, {Stead}, {Stephens}, {Tamura}, {Tappert},
  {Thompson}, {Valenti}, {Vanzi}, {Walton}, {Weidmann}, \&
  {Zijlstra}}]{Saito2012}
{Saito}, R.~K., {Hempel}, M., {Minniti}, D., {et~al.} 2012, \aap, 537, A107

\bibitem[{{Schlafly} \& {Finkbeiner}(2011)}]{Schlafly2011}
{Schlafly}, E.~F. \& {Finkbeiner}, D.~P. 2011, \apj, 737, 103

\bibitem[{{Searle} \& {Zinn}(1978)}]{SZ1978}
{Searle}, L. \& {Zinn}, R. 1978, \apj, 225, 357

\bibitem[{{Siegel} {et~al.}(2011){Siegel}, {Majewski}, {Law}, {Sarajedini},
  {Dotter}, {Mar{\'\i}n-Franch}, {Chaboyer}, {Anderson}, {Aparicio}, {Bedin},
  {Hempel}, {Milone}, {Paust}, {Piotto}, {Reid}, \& {Rosenberg}}]{Siegel2011}
{Siegel}, M.~H., {Majewski}, S.~R., {Law}, D.~R., {et~al.} 2011, \apj, 743, 20

\bibitem[{{Skrutskie} {et~al.}(2006){Skrutskie}, {Cutri}, {Stiening},
  {Weinberg}, {Schneider}, {Carpenter}, {Beichman}, {Capps}, {Chester},
  {Elias}, {Huchra}, {Liebert}, {Lonsdale}, {Monet}, {Price}, {Seitzer},
  {Jarrett}, {Kirkpatrick}, {Gizis}, {Howard}, {Evans}, {Fowler}, {Fullmer},
  {Hurt}, {Light}, {Kopan}, {Marsh}, {McCallon}, {Tam}, {Van Dyk}, \&
  {Wheelock}}]{Skrutskie2006}
{Skrutskie}, M.~F., {Cutri}, R.~M., {Stiening}, R., {et~al.} 2006, \aj, 131,
  1163

\bibitem[{{Suntzeff} {et~al.}(1991){Suntzeff}, {Kinman}, \&
  {Kraft}}]{Suntzeff1991}
{Suntzeff}, N.~B., {Kinman}, T.~D., \& {Kraft}, R.~P. 1991, \apj, 367, 528

\bibitem[{{Valenti} {et~al.}(2004){Valenti}, {Ferraro}, \&
  {Origlia}}]{Valenti2004}
{Valenti}, E., {Ferraro}, F.~R., \& {Origlia}, L. 2004, \mnras, 351, 1204

\bibitem[{{Valenti} {et~al.}(2007){Valenti}, {Ferraro}, \&
  {Origlia}}]{Valenti2007}
{Valenti}, E., {Ferraro}, F.~R., \& {Origlia}, L. 2007, \aj, 133, 1287

\bibitem[{{Valenti} {et~al.}(2010){Valenti}, {Ferraro}, \&
  {Origlia}}]{Valenti2010}
{Valenti}, E., {Ferraro}, F.~R., \& {Origlia}, L. 2010, \mnras, 402, 1729

\bibitem[{{Valenti} {et~al.}(2005){Valenti}, {Origlia}, \&
  {Ferraro}}]{Valenti2005}
{Valenti}, E., {Origlia}, L., \& {Ferraro}, F.~R. 2005, \mnras, 361, 272

\bibitem[{{Vasiliev} \& {Baumgardt}(2021)}]{Vasiliev2021b}
{Vasiliev}, E. \& {Baumgardt}, H. 2021, \mnras, 505, 5978

\bibitem[{{Vasiliev} \& {Belokurov}(2020)}]{Vasiliev2020}
{Vasiliev}, E. \& {Belokurov}, V. 2020, \mnras, 497, 4162

\bibitem[{Vasiliev {et~al.}(2021)Vasiliev, Belokurov, \& Erkal}]{Vasiliev2021a}
Vasiliev, E., Belokurov, V., \& Erkal, D. 2021, MNRAS, 501, 2279

\bibitem[{{Wagner-Kaiser} {et~al.}(2017){Wagner-Kaiser}, {Mackey},
  {Sarajedini}, {Chaboyer}, {Cohen}, {Yang}, {Cummings}, {Geisler}, \&
  {Grocholski}}]{WagnerKaiser2017}
{Wagner-Kaiser}, R., {Mackey}, D., {Sarajedini}, A., {et~al.} 2017, \mnras,
  471, 3347

\bibitem[{Wang {et~al.}(2014)Wang, Ma, Wu, \& Zhou}]{Wang_2014}
Wang, S., Ma, J., Wu, Z., \& Zhou, X. 2014, The Astronomical Journal, 148, 4

\bibitem[{White \& Rees(1978)}]{WR1978}
White, S. D.~M. \& Rees, M.~J. 1978, MNRAS, 183, 341

\end{thebibliography}

\begin{appendix}
\section{LFs and CMDs for the new Sgr GCs}
\begin{figure}[!htb]
\centering
\onecolumn
\includegraphics[width=4cm, height=4cm]{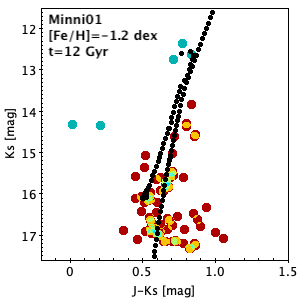}
\includegraphics[width=4cm, height=4cm]{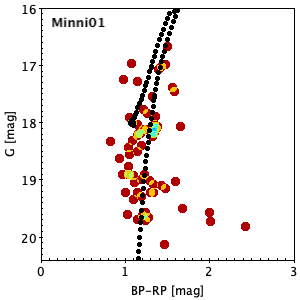}
\includegraphics[width=4cm, height=4cm]{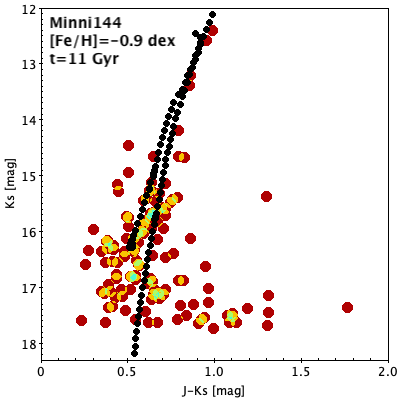}
\includegraphics[width=4cm, height=4cm]{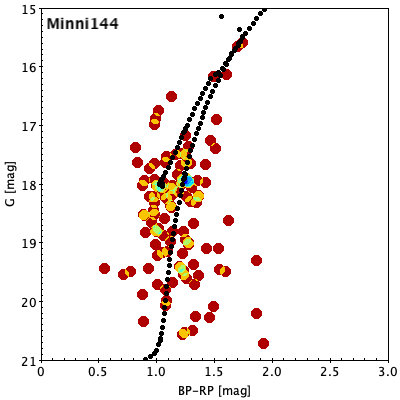}
\includegraphics[width=4cm, height=4cm]{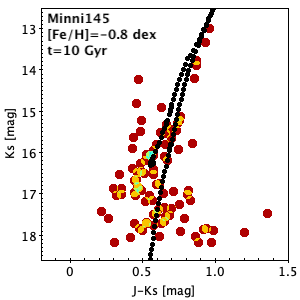}
\includegraphics[width=4cm, height=4cm]{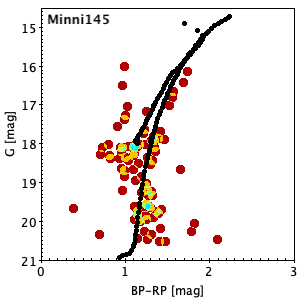}
\includegraphics[width=4cm, height=4cm]{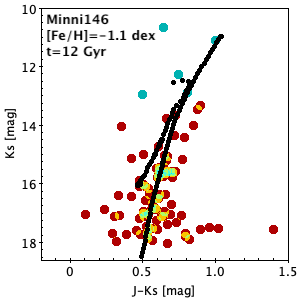}
\includegraphics[width=4cm, height=4cm]{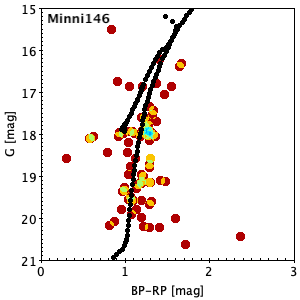}
\includegraphics[width=4cm, height=4cm]{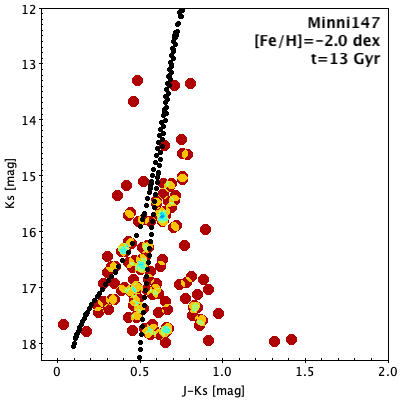}
\includegraphics[width=4cm, height=4cm]{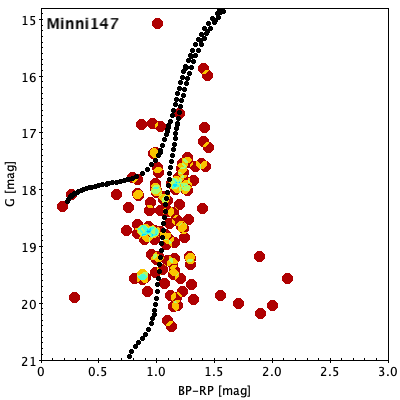}
\includegraphics[width=4cm, height=4cm]{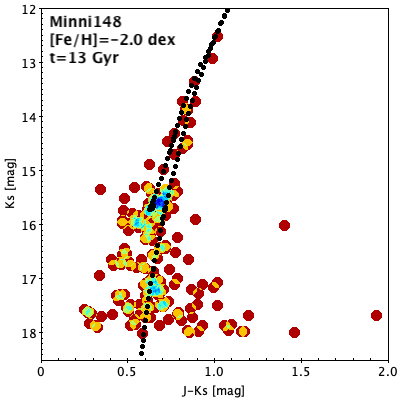}
\includegraphics[width=4cm, height=4cm]{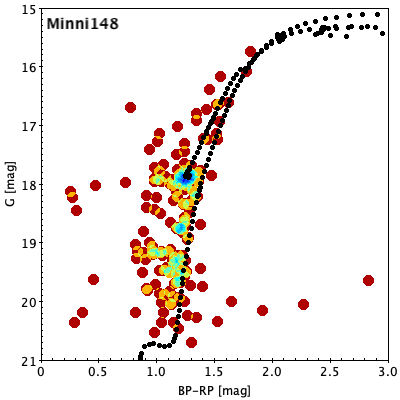}
\includegraphics[width=4cm, height=4cm]{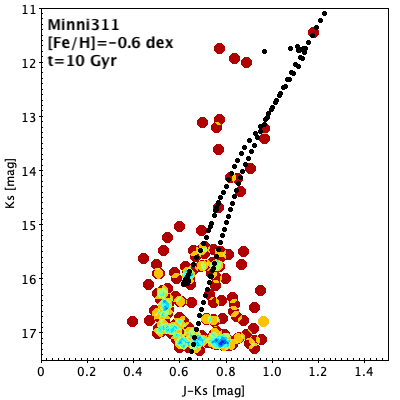}
\includegraphics[width=4cm, height=4cm]{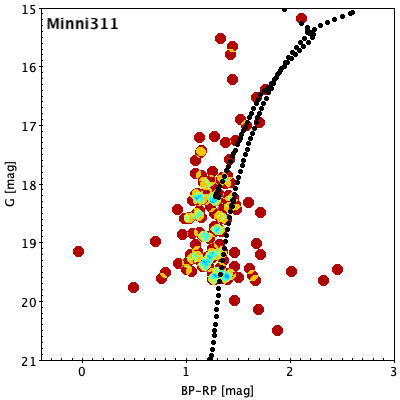}
\includegraphics[width=4cm, height=4cm]{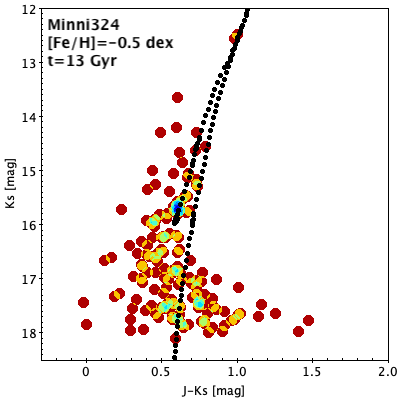}
\includegraphics[width=4cm, height=4cm]{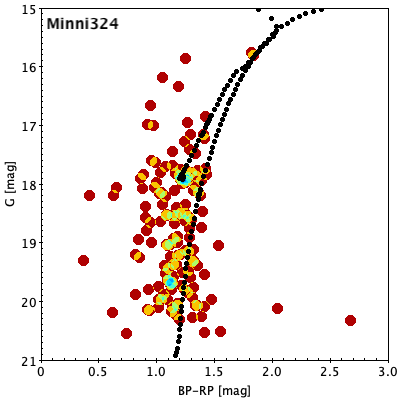}
\includegraphics[width=4cm, height=4cm]{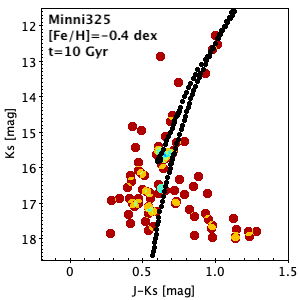}
\includegraphics[width=4cm, height=4cm]{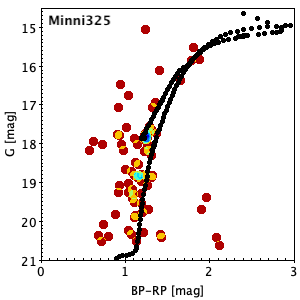}
\includegraphics[width=4cm, height=4cm]{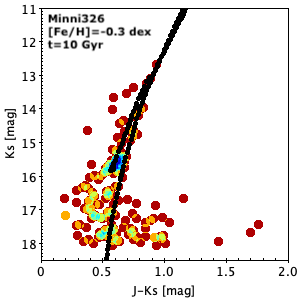}
\includegraphics[width=4cm, height=4cm]{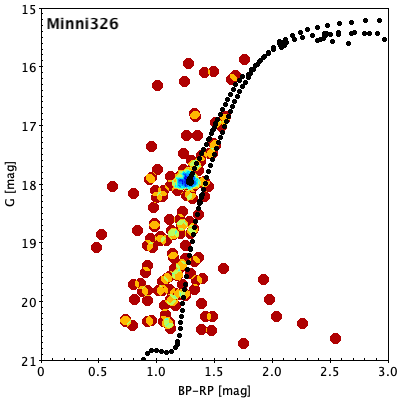}
\caption{Near-IR and optical CMDs for all the Sgr GC in our sample.  The Hess diagrams represent the VVVX (on the left) and \textit{Gaia} EDR3 (on the right) datasets, the cyan points are the stars from 2MASS catalogue,  and the black dotted lines are the PARSEC isochrones which best fit the evolutionary sequences. }
\label{cmds}
\end{figure}

\begin{figure}[!htb]
\ContinuedFloat
\captionsetup{list=off, format=continued}
\centering
\onecolumn
\includegraphics[width=4cm, height=4cm]{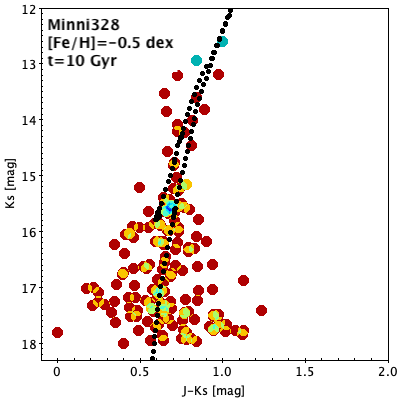}
\includegraphics[width=4cm, height=4cm]{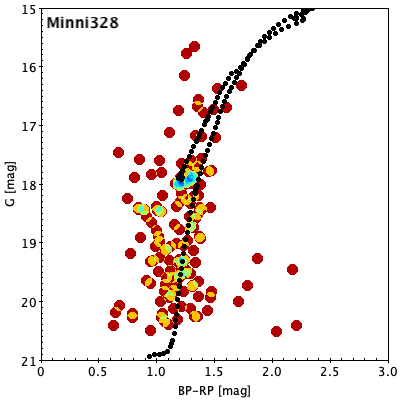}
\includegraphics[width=4cm, height=4cm]{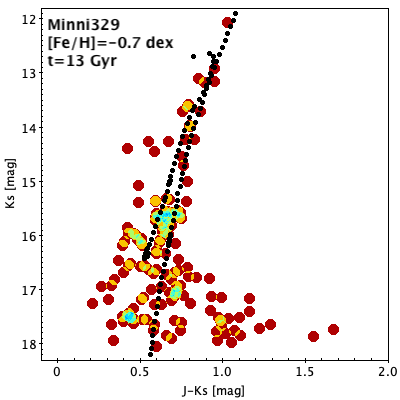}
\includegraphics[width=4cm, height=4cm]{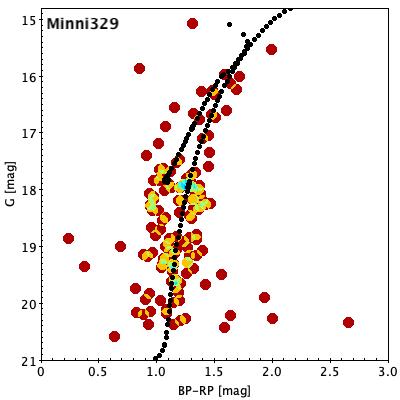}
\includegraphics[width=4cm, height=4cm]{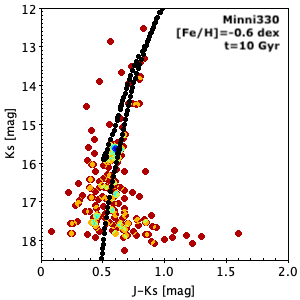}
\includegraphics[width=4cm, height=4cm]{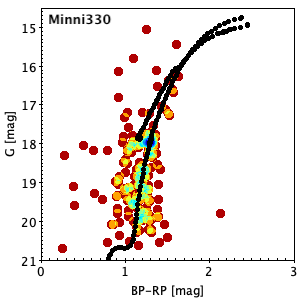}
\includegraphics[width=4cm, height=4cm]{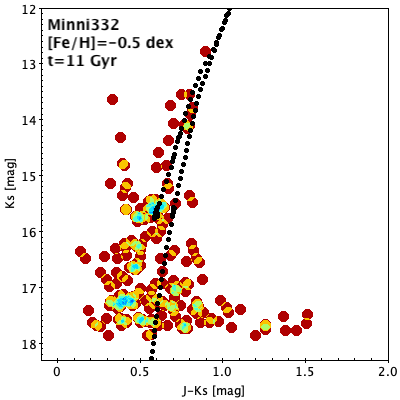}
\includegraphics[width=4cm, height=4cm]{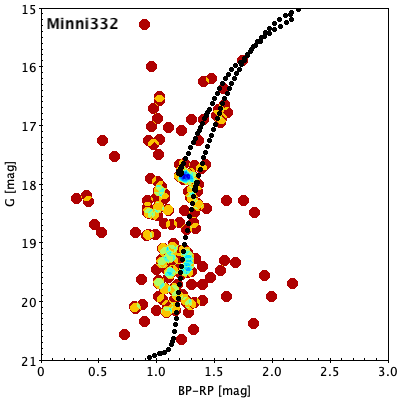}
\includegraphics[width=4cm, height=4cm]{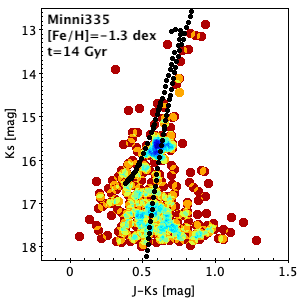}
\includegraphics[width=4cm, height=4cm]{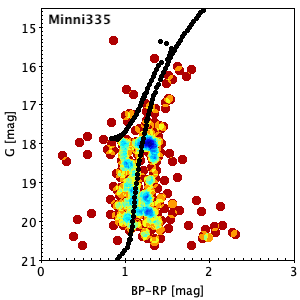}
\includegraphics[width=4cm, height=4cm]{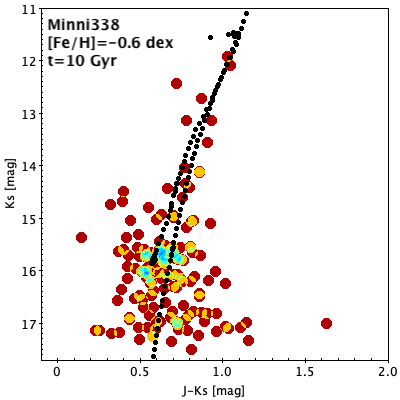}
\includegraphics[width=4cm, height=4cm]{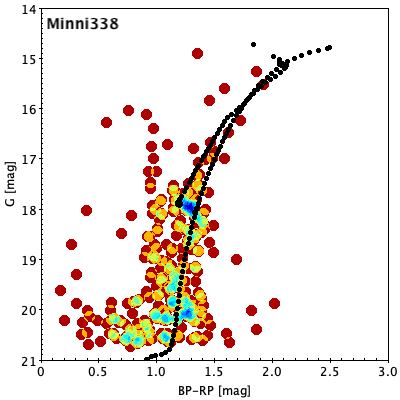}
\includegraphics[width=4cm, height=4cm]{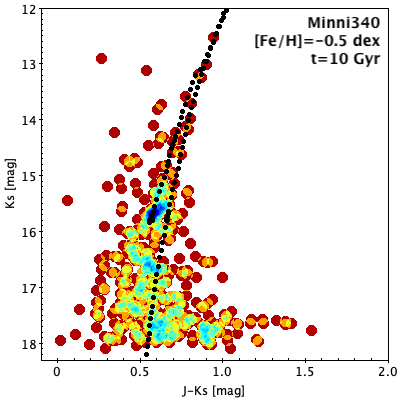}
\includegraphics[width=4cm, height=4cm]{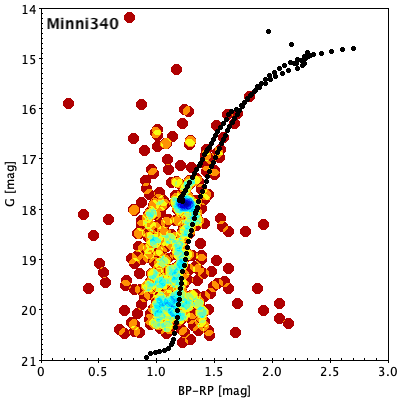}
\includegraphics[width=4cm, height=4cm]{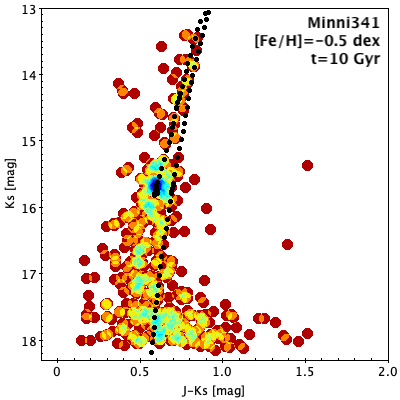}
\includegraphics[width=4cm, height=4cm]{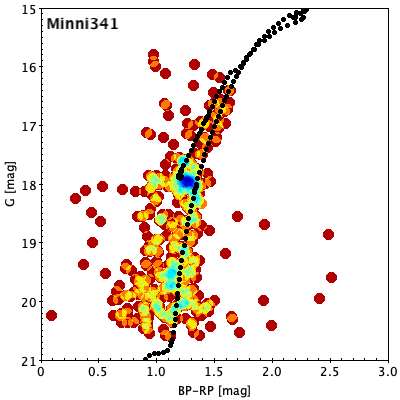}
\includegraphics[width=4cm, height=4cm]{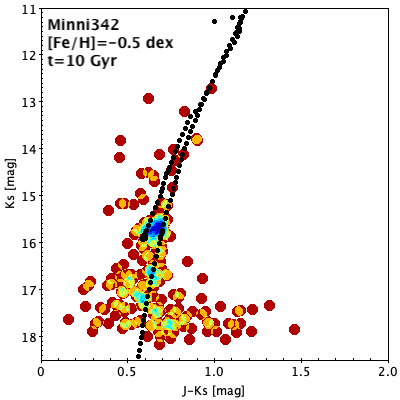}
\includegraphics[width=4cm, height=4cm]{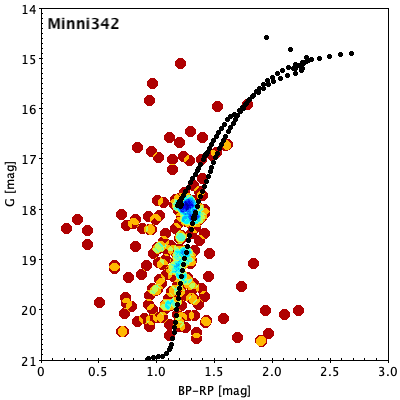}
\includegraphics[width=4cm, height=4cm]{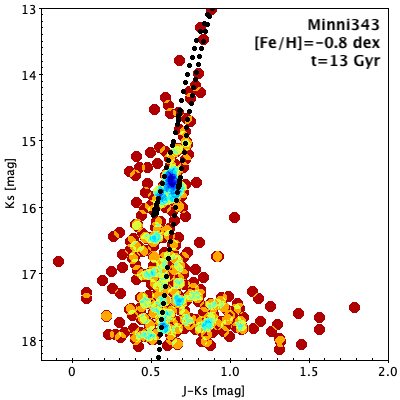}
\includegraphics[width=4cm, height=4cm]{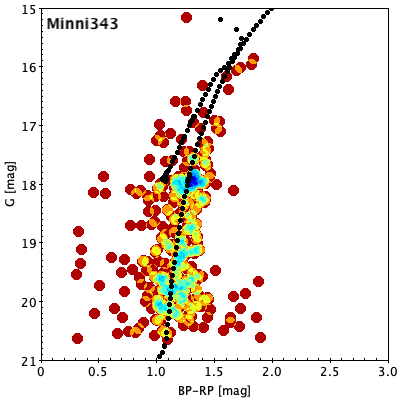}
\includegraphics[width=4cm, height=4cm]{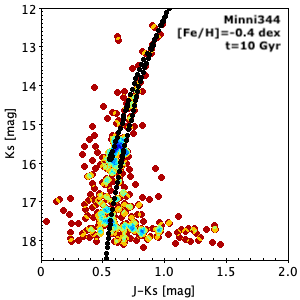}
\includegraphics[width=4cm, height=4cm]{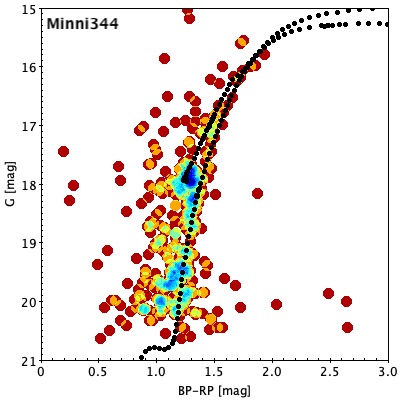}
\caption{}
\end{figure}

\begin{figure}[!htb]
\centering
\onecolumn
\includegraphics[width=6cm, height=4.5cm]{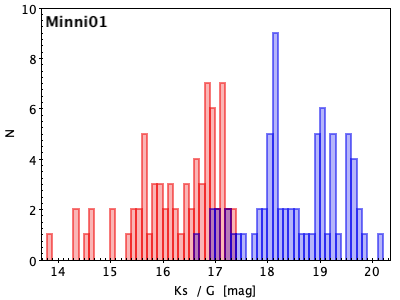}
\includegraphics[width=6cm, height=4.5cm]{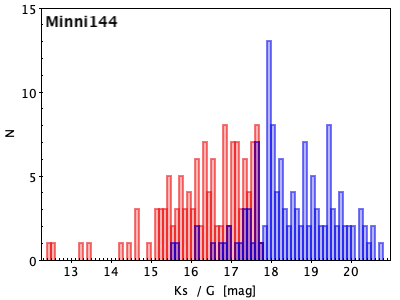}
\includegraphics[width=6cm, height=4.5cm]{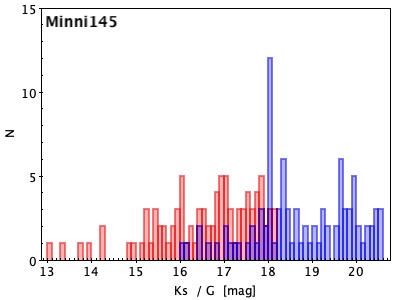}
\includegraphics[width=6cm, height=4.5cm]{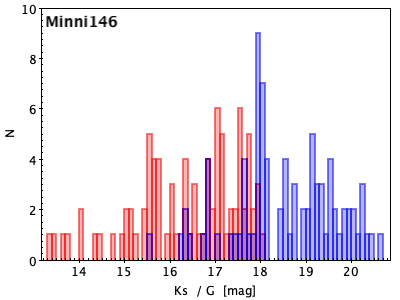}
\includegraphics[width=6cm, height=4.5cm]{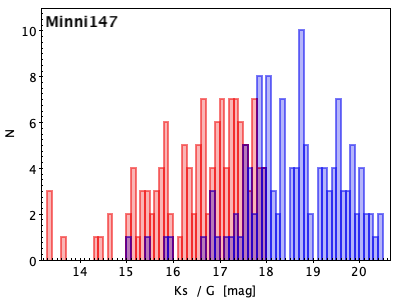}
\includegraphics[width=6cm, height=4.5cm]{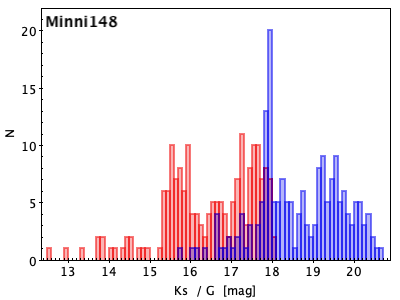}
\includegraphics[width=6cm, height=4.5cm]{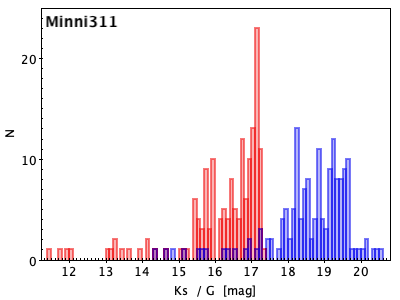}
\includegraphics[width=6cm, height=4.5cm]{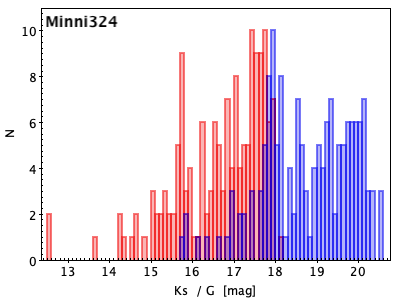}
\includegraphics[width=6cm, height=4.5cm]{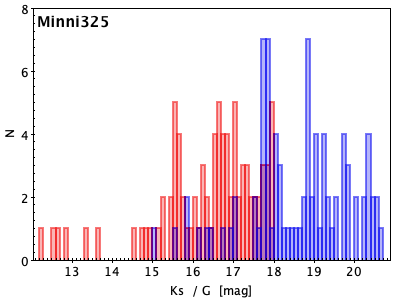}
\includegraphics[width=6cm, height=4.5cm]{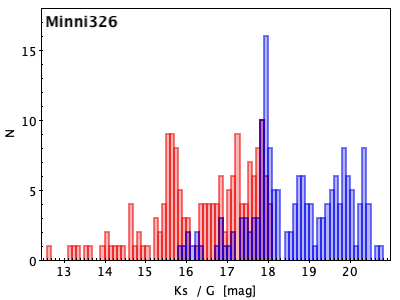}
\includegraphics[width=6cm, height=4.5cm]{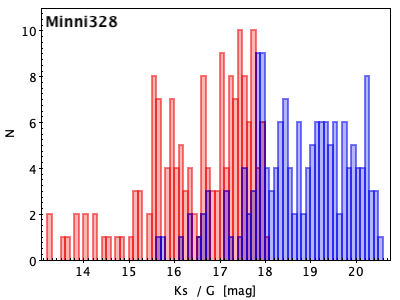}
\includegraphics[width=6cm, height=4.5cm]{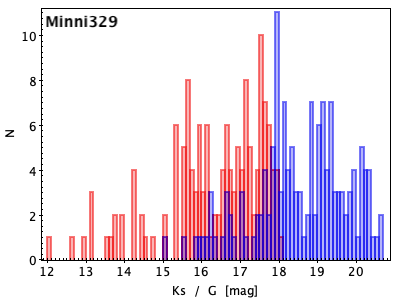}
\includegraphics[width=6cm, height=4.5cm]{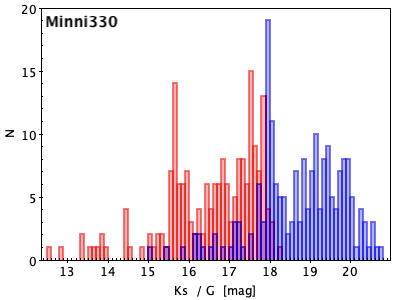}
\includegraphics[width=6cm, height=4.5cm]{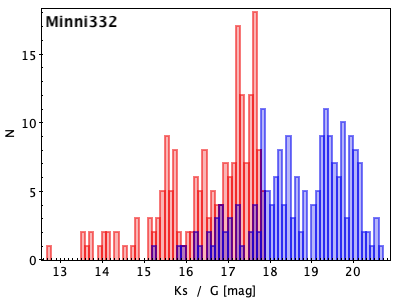}
\includegraphics[width=6cm, height=4.5cm]{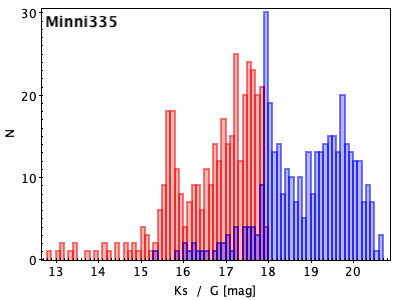}
\caption{The red and blue histograms depict the luminosity function for each Sgr GC in near-IR VVVX and in optical \textit{Gaia} EDR3 passbands, respectively.  The histograms are constructed adopting a bin size of 0.1 mag, as we want to individuate the RC position as a clear excess. }
\label{LF}
\end{figure}

\begin{figure}[!htb]
\ContinuedFloat
\captionsetup{list=off, format=continued}
\centering
\onecolumn
\includegraphics[width=6cm, height=4.5cm]{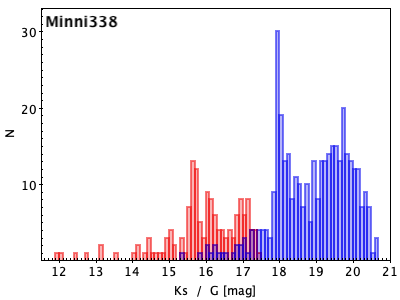}
\includegraphics[width=6cm, height=4.5cm]{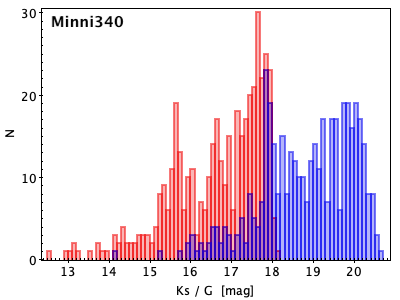}
\includegraphics[width=6cm, height=4.5cm]{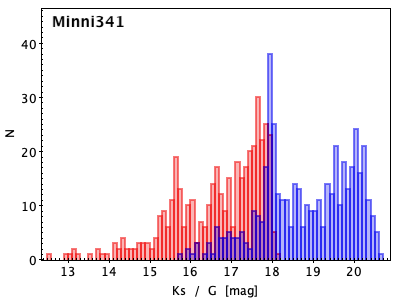}
\includegraphics[width=6cm, height=4.5cm]{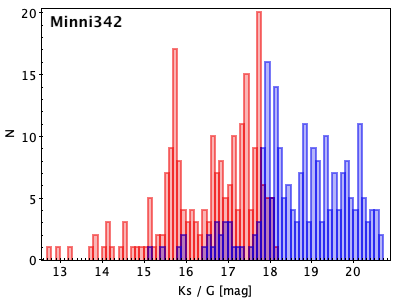}
\includegraphics[width=6cm, height=4.5cm]{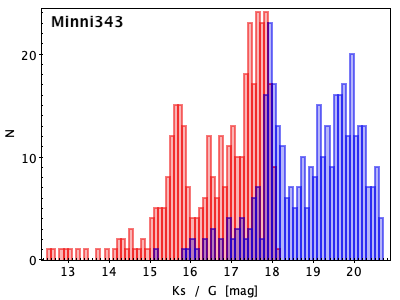}
\includegraphics[width=6cm, height=4.5cm]{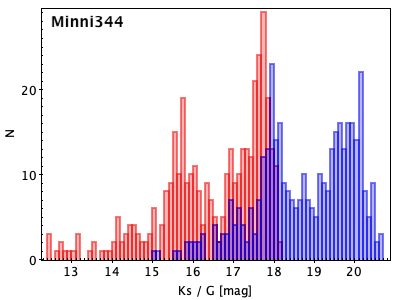}
\caption{}
\label{lf2}
\end{figure}

\twocolumn

\section{Estimation of the total luminosity}
As explained in Section 3,  the integrated luminosities of the Sgr clusters, calculated in $K_s-$band, do not include the contribution from faint stars, undetected in our photometry,  as we can appreciate in Fig. \ref{cmds}.  In order to better estimate the total luminosities of our clusters, we have therefore used the following approach.  \\

The main goal of this approach is to quantify the fraction of luminosity that comes from the faintest stars.  To do this, we compare the brightness of Sgr GCs with known and well-characterized GCs. We recover for the latter both reddening and distance modulus, listed in Table \ref{tablelum}.  The main steps adopted to achieve our intent are:
\begin{enumerate}
\item  we calculate the absolute magnitude $M_{Ks}$ of known MW GCs (Table \ref{tablelum}), using the VVV/VVVX datasets. In this way, we can directly compare the luminosities found for Sgr GCs and known GCs in the same magnitude range ($10.2 \lesssim K_s  \lesssim 17.2$,   only for Terzan 8 we used 2MASS photometry $12.0 \lesssim K_s  \lesssim 16.7$);
\item the obtained $M_{Ks}$ is converted into the absolute magnitude in $V-$band ($M_V$), using the typical GC colour $(V-K_s)=2.5$ mag (e.g.,  \citealt{Barmby2000,Cohen2007,Conroy_2010});
\item we scaled the $M_V$ values to the absolute magnitude listed in the 2010 version of the \cite{Harris1996} catalogue ($M_{V}^{Harris}$).  Consequently, we estimate the fraction of luminosity that comes from the faintest stars for each known GC, computing $\Delta(M_V)=M_{V}^{Harris}-M_V$;
\item we group these GCs according to their metallicities, then we calculate the average of $\Delta(M_V)$ fractions for each group;
\item finally, we add these averages to the Sgr luminosities. We have considered at least two known GCs with the same metallicity range as the Sgr GCs (from $-0.3$ to $-2.0$). 
\end{enumerate}

We have therefore achieved an empirical correction to our cluster's luminosities under the assumption of similarity with other GCs. The resulting luminosities are listed in Table \ref{tablegcs}. We find that the missing luminosity was smaller than 1.8 mag.

\begin{table}
\onecolumn
\centering 

\begin{tabular}{lcccl}
\hline\hline
Cluster ID &    $[Fe/H]$\tablefootmark{a}  &    $E(B-V)$ & $(m-M)_0$   &  References \\
                  & [dex] & [mag] & [mag] & \\
\hline
Liller 1 & $-0.33$ & 3.09 & 14.48 &\cite{Valenti2010} (V10)  \\ 
NGC 6440 & $-0.36$ &1.15 & 14.58&\cite{Valenti2004} (V04)  \\
NGC 6441 & $-0.46$ & 0.52 & 15.65& V04 \\
NGC 6624 & $-0.44$ &0.31  &17.33 & \cite{Siegel2011} (S11)\\
Terzan 6 & $-0.56$ & 2.35 & 14.13&\cite{Valenti2007} (V07)  \\
Terzan 12 & $-0.50$ &2.06 & 12.65 & \cite{Ortolani1998}\\
NGC 6637 & $-0.64$ & 0.22  & 17.35& S11\\
Terzan 2 & $-0.69$ &1.40 & 14.30&\cite{Christian1992}  \\
BH 261 & $-0.76$ &0.36 &13.90 & \cite{Ortolani2006}\\
NGC 6569 & $-0.76$ &0.49 &15.40 & V07 \\
UKS 1 & $-0.98$ &2.2 & 16.01& \cite{Minniti2011}\\
NGC 6638 & $-0.95$ & 0.43&15.07 &V07\\
Terzan 9 & $-1.05$ & 1.79 &13.73 & V10\\
NGC 6642 & $-1.26$&0.42 &14.30 & \cite{Barbuy2006} \\
NGC 6626 & $-1.32$ & 0.42 &13.70 & \cite{Kerber2018}\\
NGC 6540 & $-1.35$ &0.66 &13.57 & V10 \\
NGC 6558 & $-1.32$ &0.50  & 14.59& \cite{Barbuy2018} \\
NGC 6453 & $-1.50$ & 0.69 & 15.15 & V10\\
NGC 6715 & $-1.49$ &0.14 &17.27 &S11\\
Terzan 8 & $-2.16$ & 0.14 & 17.26 &S11 \\
\hline\hline
\end{tabular}
\caption{Known and well-characterised GCs used to derive the total luminosity for each Sgr GCs.}
\tablefoot{
   \tablefoottext{a}{The [Fe/H] values are taken from the 2010 version of the \cite{Harris1996} catalogue.} }
\label{tablelum}
\end{table}

\end{appendix}

\end{document}